\documentclass[aps,prb,reprint]{revtex4-2}
\usepackage{graphicx}% Include figure files
\usepackage{latexsym}
\usepackage{dcolumn}% Align table columns on decimal point
\usepackage[utf8]{inputenc}
\usepackage[T1]{fontenc}
\usepackage{tabularx}
\usepackage{hyperref}
\usepackage{color}
\usepackage{placeins}
\usepackage{float}
\usepackage{amsmath}
\usepackage{gensymb}

\begin{document}

\title{Mechanisms of radiation-induced structural transformations in deposited gold clusters}

\author{Alexey V. Verkhovtsev}
\email{verkhovtsev@mbnexplorer.com}
\affiliation{MBN Research Center, Altenh\"oferallee 3, 60438 Frankfurt am Main, Germany}
\author{Yury Erofeev}
\altaffiliation{Present address: Faculty of Geoscience, Utrecht University, Princetonlaan 8a, 3584 CB Utrecht, The Netherlands}
\affiliation{MBN Research Center, Altenh\"oferallee 3, 60438 Frankfurt am Main, Germany}
\author{Andrey V. Solov'yov}
\affiliation{MBN Research Center, Altenh\"oferallee 3, 60438 Frankfurt am Main, Germany}

%\date{\today}

\begin{abstract}
Physical mechanisms of structural transformations in deposited metallic clusters exposed to an electron beam of a transmission electron microscope (TEM) are studied theoretically and computationally. Recent TEM experiments with size-selected Au$_{923}$ clusters softly deposited on a carbon substrate showed that the clusters undergo structural transformations from icosahedron to decahedron and face-center cubic (fcc) structures upon exposure to a 200-keV electron beam.
In this paper, we demonstrate that the relaxation of collective electronic (plasmon) excitations formed in deposited metal clusters can induce the experimentally observed structural transformations. Such excitations in the clusters are formed mainly due to the interaction with low-energy secondary electrons emitted from a substrate. The characteristic occurrence times for plasmon-induced energy relaxation events are several orders of magnitude shorter than those for the momentum transfer events by energetic primary electrons to atoms of the cluster. The theoretical analysis is complemented by molecular dynamics simulations, which show that an icosahedral Au$_{923}$ cluster softly deposited on graphite is transformed into an fcc-like structure due to the vibrational excitation of the cluster.
\end{abstract}

\maketitle

%%%%%%%%%%%%%%%%%%%%%%%%%%%%%%%%%%%%%%%%%%%%%%%%%%%%%%%%%%%%%%%%%%%%%
%%%%%%%%%%%%%%%%%%%%%%%%%%%%%%%%%%%%%%%%%%%%%%%%%%%%%%%%%%%%%%%%%%%%%
\section{Introduction}

The investigation of structural transformations in atomic clusters has attracted the strong interest of the atomic cluster community for a long time \cite{Haberland_clusters_book, Meiwes-Broer_Clusters_book, LatestAdvances_2008_book, FrontiersNanoscience_vol12}.
Particular focus has been made on the study of phase transitions in atomic clusters
\cite{Haberland_clusters_book, Ekardt_MetalClusters_book, Ferrando_book, Ferrando_RMP_2005, Barnard_RepProgPhys_2010, DySoN_book_Springer_2022}, the evaluation of the melting temperature of clusters and its relation to the melting temperature of corresponding bulk materials \cite{Schmidt_1998_Nature.393.238, Haberland_2005_PRL.94.035701, Aguado_2011_AnnuRevPhysChem.62.151}, and the analysis of cluster transformations due to the relaxation of electronic excitations into the vibrational degrees of freedom \cite{Solov'yov_2005_IJMPB.19.4143, Saalmann_1998_PRL.80.3213, Gerhardt_2003_CPL.3.454}.

Another hot research topic has been related to studying the structure and dynamics of clusters deposited on surfaces using high-resolution transmission electron microscopy (TEM) and scanning transmission electron microscopy (STEM) \cite{Smith_1986_Science.233.872, Iijima_1986_PRL.56.616, Marks_1994_RepProgPhys.57.603, Li_2008_Nature.451.46, Wang_2012_PRL.108.245502, Plant_2014_JACS.136.7559, Foster_2019_NatComms.10.2583, Li_2020_ScienceAdv.6.eaay4289}.
Structural transformations in deposited nanometer-sized gold clusters exposed to intense electron-beam irradiation were observed for the first time in Ref.~\cite{Iijima_1986_PRL.56.616}. In that study, the size of deposited clusters varied from $\sim$1 to 10~nm as initially small clusters (below 1~nm in size) aggregated into larger structures under exposure to an electron beam.
More recently, structural transformations in isolated size-selected gold clusters exposed to an electron beam were demonstrated using STEM \cite{Wang_2012_PRL.108.245502, Plant_2014_JACS.136.7559}.
A series of experimental studies demonstrated that deposited gold clusters of a specific size, e.g. Au$_{20}$ \cite{Li_2020_ScienceAdv.6.eaay4289, Wang_2012_Nanoscale.4.4947}, Au$_{55}$ \cite{Wang_2012_NanoLett.12.5510}, Au$_{309}$ \cite{Li_2008_Nature.451.46}, and Au$_{923}$ \cite{Wang_2012_PRL.108.245502, Plant_2014_JACS.136.7559}, exhibit different atomic configurations which have been visualized by means of STEM.

In Ref.~\cite{Wang_2012_PRL.108.245502}, size-selected gold clusters containing $923 \pm 23$ atoms were soft-landed on a carbon substrate and irradiated by a 200-keV electron beam of a STEM. The effect of electron irradiation on the atomic structure of Au$_{923}$ clusters was studied by collecting a sequence of STEM images for $\sim$10$^2$ individual clusters.
The structure of each cluster was assigned to the icosahedron ($I_h$), decahedron ($D_h$), face-centered cubic (fcc) / octahedron ($O_h$), or amorphous/unidentified structures \cite{Wang_2012_PRL.108.245502}. It was shown that Au$_{923}$-$I_h$ isomers are unstable upon irradiation, contrary to $D_h$ and fcc Au$_{923}$ clusters that retained their structure during 400~seconds of irradiation by the electron beam.
Monitoring beam-induced structural transformations in the electron microscope revealed that most $I_h$ clusters had been converted into $D_h$ or fcc isomers upon exposure to electron beam irradiation, and no further structural transformations have been observed after the $I_h \to D_h$ or $I_h \to O_h$ transformation occurred.

The electron beam-induced dynamics of atoms in deposited metal clusters was studied computationally in Ref.~\cite{Knez_2018_Ultramicroscopy}, focusing on atomic displacement and sputtering effects. However, the physical mechanisms of electron beam-induced structural transformations in deposited metallic clusters \cite{Wang_2012_PRL.108.245502, Plant_2014_JACS.136.7559} have not been fully understood so far and require further investigation.

This study is devoted to the analysis of the physical mechanisms contributing to electron beam-induced transformations in deposited metallic clusters.
Two mechanisms of energy transfer into the deposited clusters are considered, namely elastic scattering of fast projectile electrons from cluster atoms (without excitation of the electronic subsystem of the cluster) and an inelastic scattering mechanism due to the relaxation of plasmon-type collective electronic excitations formed in the clusters.
We demonstrate that the relaxation of collective electronic excitations through the vibrational excitation of cluster atoms is a plausible mechanism for the experimentally observed structural transformations in deposited gold clusters irradiated with an electron beam.
The theoretical description of the coupling of collective electronic excitations to the vibrational modes of the ionic subsystem in free metallic clusters was given in Ref.~\cite{Gerchikov2000}. To the best of our knowledge, this phenomenon has not been discussed in connection to electron-induced structural transformations of deposited metal clusters.
The plasmon excitations in the deposited clusters are formed mainly due to the interaction with low-energy secondary electrons emitted from a substrate.
The characteristic occurrence times for plasmon-induced energy relaxation events are found to be several orders of magnitude shorter than those for the momentum transfer events, which are induced by energetic primary electrons elastically scattering from atoms of the cluster.

The theoretical analysis is complemented with classical molecular dynamics (MD) simulations performed using the MBN Explorer software package \cite{MBNExplorer_JCC_2012}. The simulations show that icosahedral Au$_{923}$ clusters softly deposited on graphite undergo a structural transformation to an fcc-like structure. This transformation is characterized by analyzing radial distribution functions and the local structural environment for different vibrationally excited cluster structures.

Physical effects occurring during the irradiation of solid specimens in TEM and STEM experiments have been widely discussed in the literature, see e.g. Refs.~\cite{Egerton_book, Jiang_2015_RepProgPhys.79.016501, Egerton_2019_Micron.119.72, Susi_2019_NatRevPhys.1.397, Egerton_2021} and references therein. Particular emphasis has been made on the analysis of radiation damage mechanisms related to knock-on atomic displacement due to the high-angle elastic scattering of fast projectile electrons and the effects caused by the inelastic scattering of these electrons.
In the case of organic and inorganic targets, energy loss by the incident primary electrons may induce radiolysis, i.e. the breakage of chemical bonds caused by electronic excitations and ionization, leading to beam-induced structural tranfsormations and degradation of the samples \cite{Jiang_2015_RepProgPhys.79.016501, Egerton_2004_Micron.35.399}.
Concerning metal systems, it was demonstrated that the focused electron beam in a TEM can split metal clusters with diameters of less than 1 nm from larger metal nanoparticles placed on carbon substrates \cite{Cretu_2012_Carbon.50.259}. The mechanism of radiation-induced fission of metal clusters might also involve collective electron excitations and their subsequent relaxation, but it can also be caused by the multiple ionization of the clusters and their subsequent decay. While the radiation-induced fission of metal clusters and nanoparticles is an interesting phenomenon, its discussion goes beyond the scope of the present study.

%%%%%%%%%%%%%%%%%%%%%%%%%%%%%%%%%%%%%%%%%%%%%%%%%%%%%%%%%%%%%%%%%%%%%
%%%%%%%%%%%%%%%%%%%%%%%%%%%%%%%%%%%%%%%%%%%%%%%%%%%%%%%%%%%%%%%%%%%%%
\section{Analysis of irradiation conditions in experiments}
\label{sec:irradiation_cond}

The theoretical analysis carried out in this study corresponds to the experimental conditions of Ref.~\cite{Wang_2012_PRL.108.245502}.
However, the analysis presented can also be generalized toward a broader range of irradiation conditions typical for TEM and STEM experiments \cite{FEI_STEM_2005, Sun_JMaterSci_STEM_2020, STEM_SmallMethods_2021} and different cluster sizes.
According to Ref.~\cite{Wang_2012_PRL.108.245502}, each cluster was irradiated for $100 - 400$ seconds at current density for the primary electron (PE) beam $j_{\textrm{PE}} = 3 \times 10^4$~$e^-$/\AA$^2$/s.
Each image series was recorded over a field of view area $S = 10.5~\textrm{nm} \times 10.5~\textrm{nm}$.
The PE beam current is then given by
\begin{eqnarray}
I = j_{\textrm{PE}} \, S &\approx& (3 \times 10^6) \, \displaystyle{ \frac{e^-}{\textrm{nm}^2 \, \textrm{s}} } \times (1.1 \times 10^2)~\textrm{nm}^2 \nonumber \\
&\approx& 5.29 \times 10^{-11} \, \displaystyle{ \frac{\textrm{C}}{\textrm{s}} } \approx 52.9~\textrm{pA} ,
\end{eqnarray}
which is a typical value of beam current used in STEM experiments \cite{Susi_2019_NatRevPhys.1.397, STEM_SmallMethods_2021}.

Radius of a Au$_{923}$ cluster can be evaluated as follows:
\begin{equation}
R = r_s \, N^{1/3} ,
\label{eq:cluster_radius}
\end{equation}
where $r_s$ is the Wigner-Seitz radius and $N$ is the number of atoms in the cluster.
For the sake of simplicity, we consider the cluster as a spherically symmetric system but the analysis performed can be generalized for non-spherical cluster geometries \cite{Kreibig_Vollmer_book}.
Using the value $r_s = 3.01~\textrm{a.u.} \approx 1.592$~\AA~for gold \cite{Cluster_Dynamics_book} and $N = 923$
one derives the cluster radius $R \approx 1.55$~nm and its cross-sectional area $S_{\textrm{cl}} = \pi R^2  \approx 7.55~\textrm{nm}^2$.

The probability of hitting a Au$_{923}$ cluster by a PE per unit time is equal to:
\begin{equation}
P_{\textrm{PE}} = j_{\textrm{PE}} \, S_{\textrm{cl}}
\approx 0.023~\displaystyle{ \textrm{ns}^{-1} } .
\label{eq:eq01}
\end{equation}
Thus, for the experimental conditions of Ref.~\cite{Wang_2012_PRL.108.245502} the average time between two subsequent hits of the Au$_{923}$ cluster by PEs is
\begin{equation}
\tau_{\textrm{PE}} = P_{\textrm{PE}}^{-1} \approx 44.2~\textrm{ns} .
\end{equation}

According to the experimental data from Ref.~\cite{Iakoubovskii_2008_PRB} and the NIST Electron Inelastic-Mean-Free-Path Database
\cite{NIST_IMFP_database, Powell_1999}, the inelastic mean free path (MFP) for a 200-keV electron in gold is $\lambda_{\rm inel} \sim 84$~nm.
Therefore, such energetic electrons will deposit their energy mainly in the substrate but not the deposited clusters of the considered size.

%%%%%%%%%%%%%%%%%%%%%%%%%%%%%%%%%%%%%%%
\begin{figure}[t!]
\centering
\includegraphics[width=0.44\textwidth]{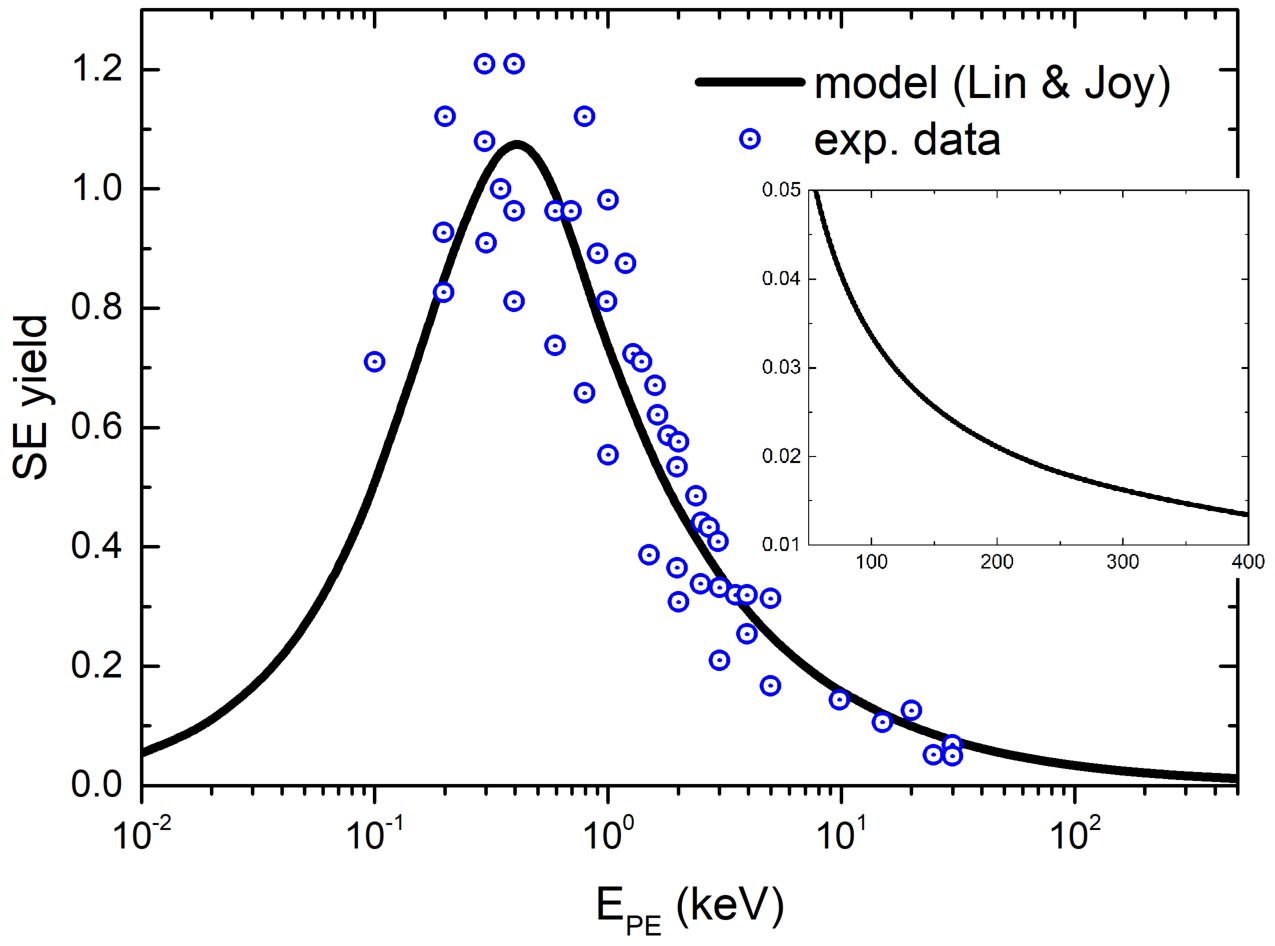}
\caption{Yield of secondary electrons (that is, the number of secondary electrons (SEs) per one primary electron (PE)) emitted from a carbon target irradiated with electrons as a function of the PE energy. The solid line shows the result obtained by means of the semi-empirical model~\cite{Lin_Joy_2005}. Symbols show experimental data for the SE yield from graphite~\cite{Bellissimo_2020_JElSpectr}. }
\label{fig:SE_yield_carbon}
\end{figure}
%%%%%%%%%%%%%%%%%%%%%%%%%%%%%%%%%%%%%%%

The interaction of the PE beam with the substrate leads to the generation of low-energy secondary electrons (SEs).
The number of SEs generated per one PE of specific energy, $N_{\textrm{SE}}(E)$ (the parameter denoted hereafter as SE yield), has been evaluated by means of a semi-empirical model described in Ref.~\cite{Lin_Joy_2005}. The model is based upon the concept of a ``universal yield curve of SE production'', where the yield of SE emission as a function of PE energy has been determined for targets made of different elements based on several parameters, namely the mass density of a target material, the effective energy required to produce an SE, and the effective SE escape depth \cite{Lin_Joy_2005}.
In the cited study, the SE yields from 44 different targets were evaluated by means of the model approach which agreed with experimental data and results of Monte Carlo simulations.
Figure~\ref{fig:SE_yield_carbon} shows the yield of SEs emitted from a carbon target as a function of PE energy. The solid line shows the result obtained by means of the model \cite{Lin_Joy_2005}. Experimental data for the SE yield from graphite \cite{Bellissimo_2020_JElSpectr} are also shown for comparison by symbols.
According to the data plotted in Fig.~\ref{fig:SE_yield_carbon} the number of SEs emitted from a carbon target per one 200-keV PE is
\begin{equation}
N_{\textrm{SE}} \approx 0.02 \, N_{\textrm{PE}} .
\label{eq:SE_yield}
\end{equation}
Then, for the experimental conditions of Ref.~\cite{Wang_2012_PRL.108.245502},
current density for the SEs emitted from the substrate and interacting with the Au$_{923}$ cluster reads as:
\begin{equation}
j_{\textrm{SE}} = \frac{ N_{\textrm{SE}} }{ N_{\textrm{PE}} } \, j_{\textrm{PE}}
\approx
6 \times 10^{4} \displaystyle{ \frac{e^-}{\textrm{nm}^2 \, \textrm{s}} } .
\label{eq:j_SE_substrate}
\end{equation}

%%%%%%%%%%%%%%%%%%%%%%%%%%%%%%%%%%%%%%%%%%%%%%%%%%%%%%%%%%%%%%
%%%%%%%%%%%%%%%%%%%%%%%%%%%%%%%%%%%%%%%%%%%%%%%%%%%%%%%%%%%%%%
\section{Energy deposition in the clusters due to plasmon excitations}
\label{sec:plasmons}

This section is devoted to the analysis of the energy transfer mechanism due to the relaxation of plasmon-type, collective electronic excitations in a deposited gold cluster.

The decay of collective electron excitations in metal clusters into single-electron and vibrational excitations is a widely studied phenomenon, see e.g. Refs.~\cite{Solov'yov_2005_IJMPB.19.4143, Kreibig_Vollmer_book, Wang_1993_CPL.205.521, Montag_1995_PRB.51.14686, Gerchikov2000, Stietz_2000_PRL.84.5644, Solov'yov_2005_IJMPB.19.4143, Kresin_2006_PRB.73.115412, Neukirch_2016_JPCC.116.15034} and references therein.
Collective excitations in the deposited metal clusters can be induced by both low-energy SEs (with the characteristic energy $E \sim 10^1 - 10^2$~eV) emitted from a substrate  and energetic PEs (with $E \sim 10^5$~eV).
Excitation of plasmons in metallic nanostructures by high energy electron beams (with $E \sim 10^4 - 10^5$~eV) has been studied in several experiments \cite{Bashevoy_2006_NanoLett.6.1113, Koh_2009_ACSNano.3.3015, Rossouw_2011_NanoLett.11.1499}; see also a review \cite{Wu_2017_ChemRev.118.2994} and references therein.
The manifestation of collective electron excitations in electron energy-loss spectra for electrons inelastically scattering from spherical metal particles and metal clusters was widely studied over several decades by means of different theoretical approaches, including hydrodynamic approach, random phase approximation, plasmon resonance approximation, and others; see Refs. \cite{Fujimoto_1968_JPhysSocJpn.25.1679, Lushnikov_1975_ZPhysB.21.357, Barberan_1985_PRB.31.6354, Ferrell_1987_PRB.35.7365, Michalewicz_1992_PRB.45.13664, Gerchikov_1997_JPB.30.4133, Gerchikov_2000_PRA.62.043201, Solov'yov_2005_IJMPB.19.4143, GdeAbajo_2010_PMP.82.209, Gildenburg_2016_PhysPlasmas.23.032120} and references therein.

In this study, the contribution of collective electronic excitations to the singly differential inelastic scattering cross section of the cluster
as a function of the energy loss $\Delta \varepsilon$ of the incident electron,
${\textrm{d}\sigma_{\textrm{pl}}/\textrm{d}\Delta \varepsilon}$, is calculated using the plasmon resonance approximation (PRA), described in Refs.~\cite{Solov'yov_2005_IJMPB.19.4143, Kreibig_Vollmer_book, Gerchikov_1997_JPB.30.4133} and references therein.
This approach is based on the fact that the dominating contribution to the inelastic scattering cross section in the vicinity
of the plasmon resonance comes from collective electron excitations, while single-particle excitations give a small contribution
\cite{Haberland_clusters_book, Kreibig_Vollmer_book, Solov'yov_2005_IJMPB.19.4143}.
%The validity of PRA has been proven in numerous studies through a good comparison of the cross sections calculated using PRA with the results of many-body quantum calculations and with experiment; see Refs.~\cite{Gerchikov_1998_PRL.81.2707, Solov'yov_2005_IJMPB.19.4143, LatestAdvances_2008_book, DySoN_book_Springer_2022} and references therein.
This approach has provided a clear physical explanation of the resonant-like structures in photoionization spectra and differential inelastic scattering cross sections of metallic clusters and carbon fullerenes irradiated by electrons and ions as well as a good agreement with the results of many-body quantum calculations and with experiment; see Refs.~\cite{Gerchikov_1998_PRL.81.2707, Solov'yov_2005_IJMPB.19.4143, LatestAdvances_2008_book, DySoN_book_Springer_2022} and references therein.

Within the framework of PRA, the cross section ${\textrm{d}\sigma_{\textrm{pl}}/\textrm{d}\Delta \varepsilon}$ reads as \cite{Gerchikov_1997_JPB.30.4133, Solov'yov_2005_IJMPB.19.4143}:
\begin{eqnarray}
\frac{ {\textrm{d}\sigma_{\textrm{pl}} }}{ \textrm{d}\Delta \varepsilon } &=& \frac{8 e^2 R^3}{v^2} \sum_l (2l+1)^2 \, S_l\left(\frac{\Delta \varepsilon \, R}{v}\right) \nonumber \\
&\times& \frac{  \omega_l^2 \Delta \varepsilon \, \Gamma_l  }{ (\Delta \varepsilon^2 - \omega_l^2)^2 + \Delta \varepsilon^2 \Gamma_l^2 } \ .
\label{eq:SDCS_pl}
\end{eqnarray}
Here
$v$ is the velocity of the projectile electron,
$R$ is the cluster radius defined by Eq.~(\ref{eq:cluster_radius}), and
\begin{equation}
    \omega_l = \sqrt{ \frac{3 l \, N_e}{(2l+1) R^3}  }
\label{eq:plasmon_freq}
\end{equation}
is the frequency of the plasmon excitation with angular momentum $l$.
The function $S_l \left(\Delta \varepsilon \, R / v \right)$ reads as
\begin{equation}
    S_l \left(\frac{\Delta \varepsilon \, R}{v}\right) =
    \int_{q_{\textrm{min}}R}^{q_{\textrm{max}}R} \frac{\textrm{d}x}{x^3} \, j_l^2(x)
\label{eq:funct_S_l}
\end{equation}
where
\begin{eqnarray}
q_{\textrm{min}} &=& p (1 - \sqrt{1 - \Delta \varepsilon/E}) \ , \nonumber \\
q_{\textrm{max}} &=& p (1 + \sqrt{1 - \Delta \varepsilon/E})
\end{eqnarray}
are the minimum and maximum values of the transferred momentum,
$p$ is the momentum of the projectile electron,
$j_l(x)$ is a spherical Bessel function of the order $l$, and
$\Delta \varepsilon \, R/v$ is a dimensionless parameter.
The calculated cross section ${\textrm{d}\sigma_{\textrm{pl}}/\textrm{d}\Delta \varepsilon}$ accounts for the contributions of plasmon excitations of multipole terms (up to $l = 3$), because the excitations with higher angular momentum are formed by single-electron transitions rather than the collective ones
\cite{Solov'yov_2005_IJMPB.19.4143}.
Explicit expressions for the function $S_l$ for different values of $l$ are given in Appendix.
The parameter $\Gamma_l$ in Eq.~(\ref{eq:SDCS_pl}) is set equal to 4~eV, following our earlier studies of photoabsorption and inelastic scattering of protons from small gold clusters and nanometer-sized gold nanoparticles \cite{Verkhovtsev_2015_PRL.114.063401, Verkhovtsev_2015_JPCC.119.11000}. In those studies, the width of the dipole mode of the plasmon-type resonance, $\Gamma_1$, was determined by comparing the photoabsorption cross section for several three-dimensional gold clusters calculated by means of the PRA and time-dependent density-functional theory. The same value of $\Gamma = 4$~eV has been used for higher multipole terms. The width of a few electronvolts is typical for the collective electronic resonances in metal clusters \cite{LatestAdvances_2008_book, Ekardt_MetalClusters_book, Kleinig_1998_EPJD.4.343, Hoevel_1993_PRB.48.18178}. For other metallic systems with delocalized valence electrons, such as carbon fullerenes, the plasmon resonances are even broader with the width of $\sim$10~eV.
The correspondence of the inelastic scattering cross sections due to plasmon excitations, calculated using the PRA \cite{Solov'yov_2005_IJMPB.19.4143, Kreibig_Vollmer_book, Gerchikov_1997_JPB.30.4133} and the random phase approximation \cite{Lushnikov_1975_ZPhysB.21.357, Lushnikov_1974_ZPhysik.270.17} was discussed previously in the review \cite{Solov'yov_2005_IJMPB.19.4143}.

%%%%%%%%%%%%%%%%%%%
\begin{figure}[t!]
\centering
\includegraphics[width=0.46\textwidth]{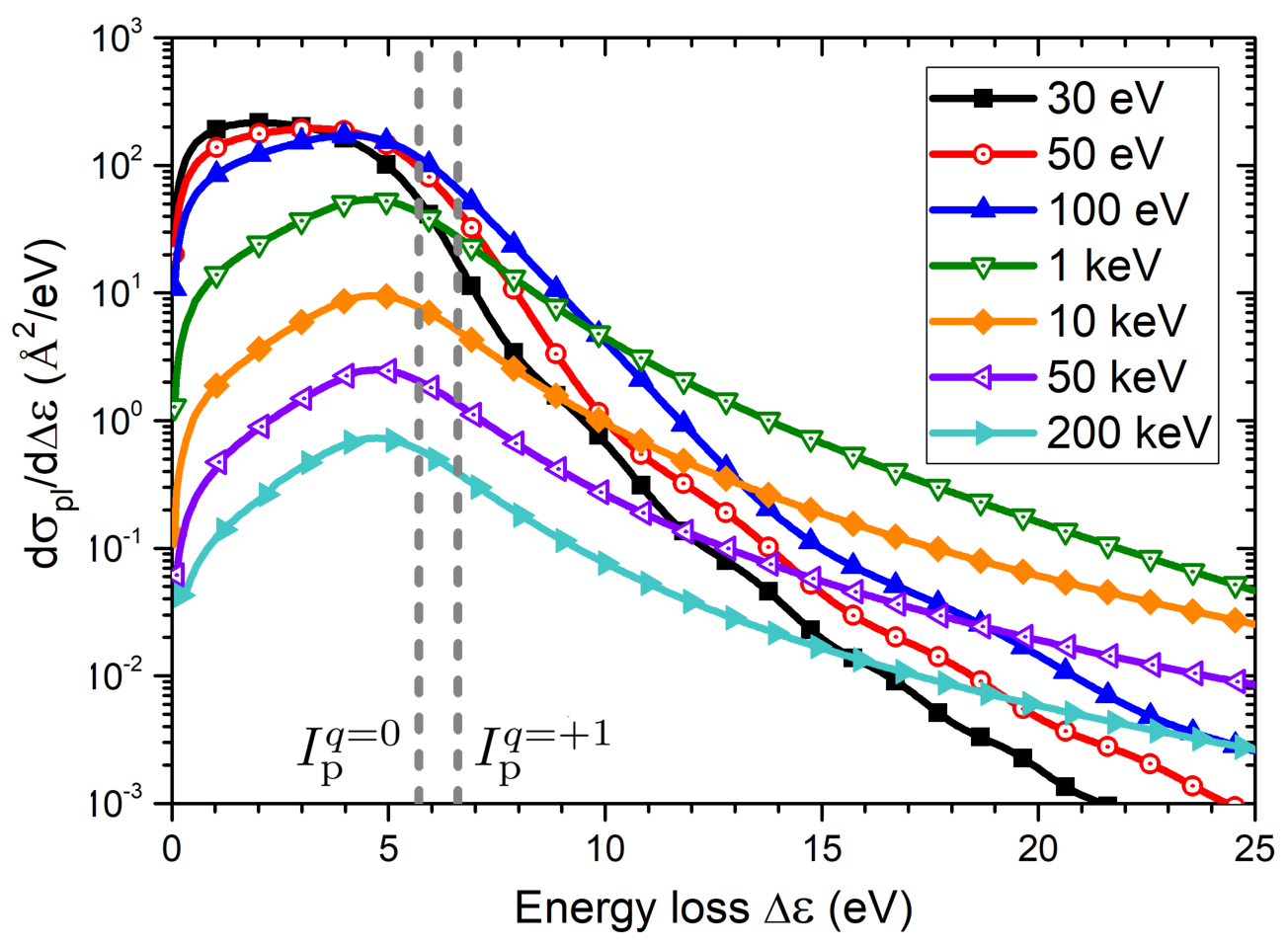}
\caption{Contribution of the plasmon excitations to the singly differential inelastic scattering cross section, ${\textrm{d}\sigma_{\textrm{pl}}/\textrm{d}\Delta \varepsilon}$, for a Au$_{923}$ cluster irradiated with $E = 30$~eV -- 200 keV electrons as a function of their energy loss $\Delta \varepsilon$. The dashed lines show the ionization potentials of the neutral and singly charged Au$_{923}$ clusters, $I_{\rm p}^{q = 0} \approx 5.7$~eV and $I_{\rm p}^{q = +1} \approx 6.6$~eV, calculated using Eq.~(\ref{eq:ionization_pot}).}
\label{fig:plasmon_CS}
\end{figure}
%%%%%%%%%%%%%%%%%%%

Figure~\ref{fig:plasmon_CS} shows the contribution of the collective electronic excitations to the cross section ${\textrm{d}\sigma_{\textrm{pl}}/\textrm{d}\Delta \varepsilon}$ for a Au$_{923}$ cluster as a function of the energy loss $\Delta \varepsilon$ of the incident electron.
As shown in the figure, the amplitude and the shape of the plasmon resonance depend on the kinetic energy of the projectile electron.
The maximum cross section for a SE with the characteristic energy of 30~eV exceeds by more than two orders of magnitude the cross section for a 200-keV PE.
The shape of the plasmon resonance varies because the relative contributions of non-dipole terms ($l = 2$ and 3) to the cross section decrease significantly with an increase of the collision velocity.

The collective electronic excitations dominate the electron energy loss spectrum at small values of $\Delta \varepsilon$ in the vicinity of the plasmon resonance frequency, while the plasmon contribution drops off at higher $\Delta \varepsilon$ values above the ionization potential $I_{\rm p}$ of the cluster.
At excitation energies $\Delta \varepsilon > I_{\rm p}$, inelastic scattering of the projectile electron results in the emission of a secondary electron. In this case, an outgoing electron carries away most of the energy transferred to the cluster by the projectile electron, and only a small fraction of the transferred energy can remain within the cluster after the electron emission.

In the case of large momentum and energy transfer events providing the main contribution to the total cross section of inelastic scattering, the electron-impact ionization cross section of the cluster can be estimated as an incoherent sum of contributions generated in binary electron--electron collisions involving the cluster atoms \cite{Korol_AVS_BrS_2014}. Therefore, the total ionization cross section of the cluster represents the incoherent sum of the ionization cross sections of $N$ individual atoms, $\sigma_{\rm ion}({\rm Au}_{923}) \approx N \sigma_{\rm ion}$, where $N = 923$.
The ionization cross section of gold for a 200-keV electron calculated using the relativistic Binary Encounter Bethe model is $\sigma_{\rm ion} \approx 0.037$~\AA$^2$ \cite{Sakata_2016_JAP.120.244901}, which gives the ionization cross section for the cluster $\sigma_{\rm ion}({\rm Au}_{923}) \sim 34.1$~\AA$^2$. For the PE current density considered in this study, $j_{\textrm{PE}} = 3 \times 10^4$~\AA$^{-2}$s$^{-1}$, the characteristic occurrence time for the ionization of the Au$_{923}$ cluster by PEs is $\tau \sim 0.98$~$\mu$s, which is significantly longer than the typical relaxation times for excited electronic states in metallic clusters (which are on a (sub)-picosecond timescale \cite{Gerchikov2000}).
%gold (which are on the $10^{-15} - 10^{-13}$~s timescale \cite{Aeschlimann_2000_ApplPhysA.71.485}).
In this paper, we do not analyze in greater detail possible effects induced by the inelastic scattering of energetic PEs, such as charge transfer effects or charge accumulation in the deposited gold cluster. These interesting problems are beyond the scope of this paper and deserve separate consideration.

%%%%%%%%%%%%%%%%%%%%%%%%%
\begin{figure}[t!]
\centering
\includegraphics[width=0.43\textwidth]{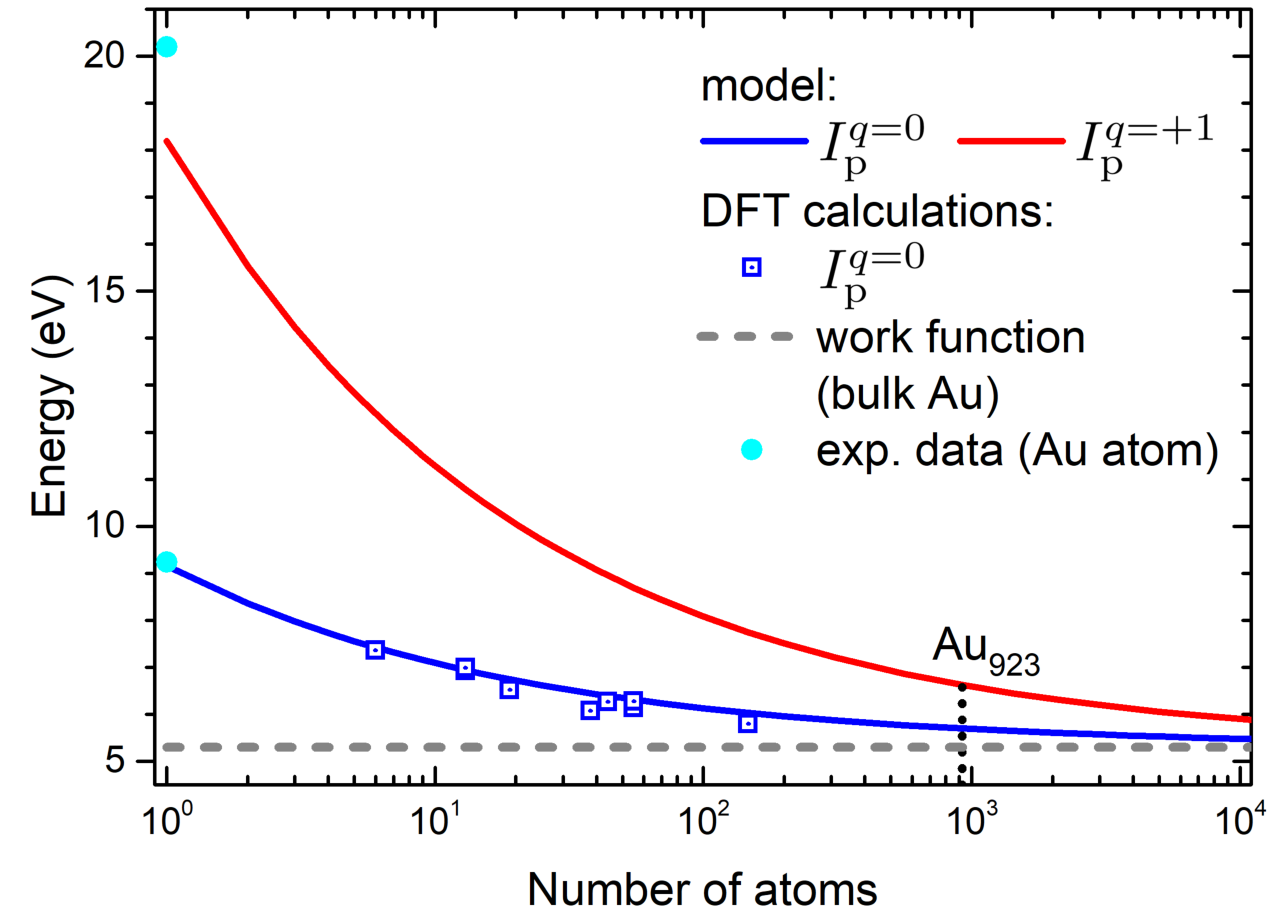}
\caption{Ionization potentials $I_{\rm p}$ for different neutral ($q=0$) and singly charged ($q=+1$) gold clusters as functions of the number of atoms $N$. The solid lines have been obtained by means of Eq.~(\ref{eq:ionization_pot}). The dashed line shows the work function of bulk gold. Solid symbols show the ionization thresholds for an Au atom and singly charged Au$^+$ ion \cite{NIST_Handbook_Atomic_Data}. Open symbols denote the results of DFT calculations \cite{Haeberlen_1997_JCP.106.5189} for Au$_N$ ($N = 6-147$) clusters. }
\label{fig:ioniz_pot_el_affinity}
\end{figure}
%%%%%%%%%%%%%%%%%%%%%%%%%

Vertical dashed lines in Fig.~\ref{fig:plasmon_CS} show the ionization thresholds for a neutral Au$_{923}$ cluster and its singly charged positive ion Au$_{923}^+$. The ionization threshold for a cluster carrying a charge $q$ has been obtained by means of a spherical jellium model according to Refs.~\cite{Haberland_clusters_book, Seidl_1991_JCP.95.1295, Seidl_1998_JCP.108.8182}:
\begin{equation}
I_{\rm p}^{q} = W + (\alpha + c)\frac{e^2}{R} + \frac{qe^2}{R} + O(R^{-2}) \ .
\label{eq:ionization_pot}
\end{equation}
Here $R$ is the cluster radius defined through the number of valence electrons in the cluster and the Wigner-Seitz radius $r_s$ (see Eq.~(\ref{eq:cluster_radius}));
$W$ is the electron work function of bulk metal;
$\alpha = 1/2$ stems from the classical model describing the metal cluster as a perfectly conducting sphere;
and the parameter $c \equiv c(r_s)$ accounts for a quantum correction due to spill-out of electron density \cite{Seidl_1996_AnnPhys.245.275}.
This parameter was determined in Ref.~\cite{Seidl_1998_JCP.108.8182} for different $r_s$ values in the range $r_s = (2 - 6)$~a.u.
For gold ($r_s \approx 3.01$~a.u.) the parameter $c \approx -0.074$.

Figure~\ref{fig:ioniz_pot_el_affinity} shows the ionization potentials for different neutral and singly charged gold clusters, Au$_N$ and Au$_N^+$, as functions of the number of atoms $N$. The solid lines have been obtained by means of Eq.~(\ref{eq:ionization_pot}).
In the limit $N \to \infty$ the ionization potentials converge to the electron work function $W$ of a bulk material. According to Ref.~\cite{CRC_Handbook_Chem-Phys}, the value of $W$ for gold varies from 5.1 to 5.5~eV. In the present study we have used the mean value $W = 5.3$~eV (see the dashed line in Fig.~\ref{fig:ioniz_pot_el_affinity}). The calculated ionization potentials of neutral and singly charged Au$_{923}$ clusters are equal to $I_{\rm p}^{q=0} \approx 5.7$~eV and $I_{\rm p}^{q = +1} \approx 6.6$~eV, respectively.
Solid symbols in Fig.~\ref{fig:ioniz_pot_el_affinity} show the ionization potentials for neutral and singly charged gold atoms, $I_{\rm p}({\rm Au}) = 9.23$~eV and $I_{\rm p}({\rm Au}^+) = 20.2$~eV \cite{NIST_Handbook_Atomic_Data}.
Open symbols denote the results of DFT calculations \cite{Haeberlen_1997_JCP.106.5189} for neutral Au$_N$ ($N = 6-147$) clusters.

%%%%%%%%%%%%%%%%%%%%%%%%%
\begin{figure}[t!]
\centering
\includegraphics[width=0.43\textwidth]{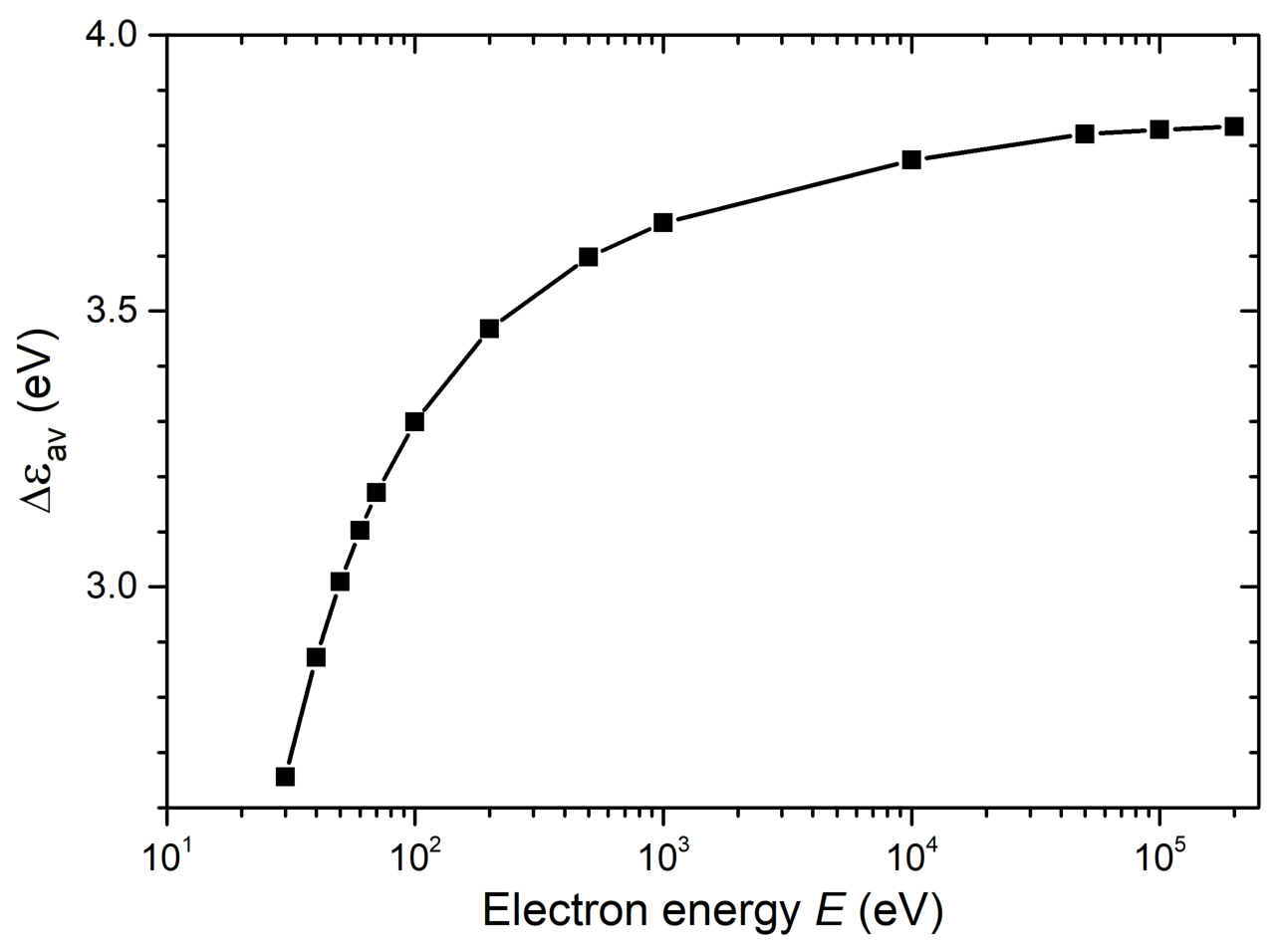}
\caption{The average amount of energy $\Delta\varepsilon_{\textrm{av}}$ transferred to the Au$_{923}$ cluster below its ionization potential by a projectile electron of a specific energy $E$, see Eq.~(\ref{eq:DeltaEps_av}). }
\label{fig:plasmon_Eloss}
\end{figure}
%%%%%%%%%%%%%%%%%%%%%%%%%

Figure~\ref{fig:plasmon_CS} shows that the maximum of the cross section ${\textrm{d}\sigma_{\textrm{pl}}/\textrm{d}\Delta \varepsilon}$ for a Au$_{923}$ cluster is located at $\Delta \varepsilon$ values below the ionization potential of the cluster. Therefore, plasmon excitations in the cluster with the excitation energies $\Delta\varepsilon < I_{\rm p}$ will decay with a significant probability through the vibrational excitation of its ionic subsystem due to the electron-phonon coupling \cite{Gerchikov2000}.

%%%%%%%%%%%%%%%%%%%%%%%%%
\begin{table*}[t!]
\centering
\caption{Characteristic appearance times for inducing a plasmon excitation in the Au$_{923}$ cluster by a 200-keV PE and a 30-eV SE in the case of transferred energy below the ionization potential of the cluster, $\tau_{\textrm{pl}}^{(1)}$, and above the ionization potential, $\tau_{\textrm{pl}}^i$. Corresponding values of the PE and SE current densities, $j_{\textrm{PE}}$ and $j_{\textrm{SE}}$, are also indicated. The numbers listed in the second row account for the additional flux density of SEs caused by the ionization of the deposited cluster by PEs, see Eqs.~(\ref{eq:j_i})--(\ref{eq:j_i_final}) and the corresponding discussion in the main text.}
\begin{tabular}{cc|cc|cc}
 \hline
       &     & \multicolumn{2}{c}{$\tau_{\textrm{pl}}^{(1)}$~($\mu$s)} & \multicolumn{2}{c}{$\tau_{\textrm{pl}}^i$~($\mu$s)} \\
    $j_{\textrm{PE}}$ $\displaystyle{ \left(\frac{e^-}{{\rm nm}^2 \, {\rm s}} \right) } $  &  $j_{\textrm{SE}}$ $\displaystyle{ \left(\frac{e^-}{{\rm nm}^2 \, {\rm s}} \right) } $   &  $E = 200$~keV  &  $E = 30$~eV  &  $E = 200$~keV  &  $E = 30$~eV  \\
 \hline
  $3 \times 10^{6}$     &   $6 \times 10^{4}$    &   14.5  &  1.8  &  15.2  &  37.9   \\
  $3 \times 10^{6}$     &   $7.3 \times 10^{4}$  &   14.5  &  1.5  &  15.2  &  31.1   \\
 \hline
\end{tabular}
\label{table:current_prob-2}
\end{table*}
%%%%%%%%%%%%%%%%%%%%%%%%%

The average excitation energy of the cluster can be calculated as
\begin{equation}
\Delta\varepsilon_{\textrm{av}} =
\frac{ \displaystyle{ \int_0^{I_{\rm p}} \Delta\varepsilon \frac{{\rm d}\sigma_{\textrm{pl}}}{{\rm d}\Delta\varepsilon} {\rm d} \Delta\varepsilon } }
{ \displaystyle{ \int_0^{I_{\rm p}} \frac{{\rm d}\sigma_{\textrm{pl}}}{{\rm d}\Delta\varepsilon} {\rm d} \Delta\varepsilon } } \ ,
\label{eq:DeltaEps_av}
\end{equation}
where the upper integration limit is set equal to
the ionization threshold of the neutral Au$_{923}$ cluster, $I_{\rm p} \approx 5.7$~eV.
The dependence of $\Delta\varepsilon_{\textrm{av}}$ on kinetic energy of the projectile electron is plotted in Fig.~\ref{fig:plasmon_Eloss}. The figure demonstrates that the average energy transferred to the Au$_{923}$ cluster below $I_{\rm p}$ due to the collision with a 30-eV secondary electron is $\Delta\varepsilon_{\textrm{av}} \sim 2.65$~eV, while a 200-keV PE will transfer to the cluster the energy $\Delta\varepsilon_{\textrm{av}} \sim 3.8$~eV.

Let us calculate the probability (per unit time) that a projectile electron (either a PE or a SE) inducing a plasmon excitation in the Au$_{923}$ cluster will transfer the amount of energy below the ionization threshold of the cluster:
\begin{equation}
P_{\rm pl}^{(1)} \equiv P_{\Delta \varepsilon \le I_{\rm p}} =
j \times
\left( { \displaystyle{ \int_0^{I_{\rm p}} \frac{{\rm d}\sigma_{\textrm{pl}}}{{\rm d}\Delta\varepsilon} {\rm d} \Delta\varepsilon } } \right) ,
\label{eq:eq02a}
\end{equation}
where $j$ is the electron current density.
The characteristic appearance time for this event reads as
\begin{equation}
\tau_{\textrm{pl}}^{(1)} = \frac{1}{ P_{\rm pl}^{(1)} } \ .
\label{eq:tau_pl_1}
\end{equation}

Table~\ref{table:current_prob-2} summarizes the values of $\tau_{\textrm{pl}}^{(1)}$ for the case of interaction with (i) a 200-keV PE and (ii) a SE with the characteristic energy of 30~eV. The values of $\tau_{\textrm{pl}}^{(1)}$ have been calculated for the experimental conditions from Ref.~\cite{Wang_2012_PRL.108.245502} and the corresponding values of $j_{\textrm{SE}}$ for SEs emitted from a carbon substrate, see the first row in Table~\ref{table:current_prob-2}.
The characteristic time $\tau_{\textrm{pl}}^{(1)}$ for a 30-eV SE, $\tau_{\textrm{pl}}^{(1)} \sim 1.8~\mu$s, is an order of magnitude shorter than for a 200-keV PE, $\tau_{\textrm{pl}}^{(1)} \sim 14.5~\mu$s.
This means that low-energy SEs emitted from the substrate will induce plasmon excitations in the cluster and transfer the amount of energy below the cluster's ionization threshold more frequently than the high-energy PEs. The number of such events occurring during the acquisition time for one STEM frame (0.8~s) reported in Ref.~\cite{Wang_2012_PRL.108.245502} is $N_{\rm pl} \sim 4.4 \times 10^5$.
Relaxation of plasmon excitations due to electron-phonon coupling will lead to an increase in the amplitude of atomic vibrations, which may initiate the experimentally observed structural transformations of the deposited gold clusters.

Now let us evaluate the probability for excitation of the second plasmon in the deposited Au$_{923}$ cluster within the period of relaxation of the first plasmon, $\tau_{\rm rel}$.
Within the time period $\tau_{\rm rel}$, the probability of a plasmon excitation is equal to
\begin{equation}
P_{\rm pl}^{(1)}(\tau_{\rm rel})
= j_{\rm SE} \, \sigma_{\rm pl} \, \tau_{\rm rel} \equiv \frac{ \tau_{\rm rel} }{ \tau_{\rm pl}^{(1)} } \ ,
\label{eq:P_pl_1}
\end{equation}
where $\sigma_{\rm pl}$ denotes the integral in Eq.~(\ref{eq:eq02a}) and $\tau_{\rm pl}^{(1)}$ is given by Eq.~(\ref{eq:tau_pl_1}).
The probability (per unit time) of excitation of the second plasmon within $\tau_{\rm rel}$ is equal to
\begin{equation}
P_{\rm pl}^{(2)}
= j_{\rm SE} \, \sigma_{\rm pl} \, P_{\rm pl}^{(1)}(\tau_{\rm rel})
 \ .
\label{eq:P_pl_2}
\end{equation}
Substituting Eq.~(\ref{eq:P_pl_1}) into (\ref{eq:P_pl_2}), one obtains
\begin{equation}
P_{\rm pl}^{(2)} = \frac{ \tau_{\rm rel} }{ \left( \tau_{\rm pl}^{(1)} \right)^2 } \ .
\label{eq:tau_pl_2-1}
\end{equation}
The characteristic appearance time for the formation of a second plasmon in the deposited Au$_{923}$ cluster within the relaxation time of the first plasmon is then equal to:
\begin{equation}
\tau_{\rm pl}^{(2)} = \frac{1}{ P_{\rm pl}^{(2)}  } \ .
\label{eq:tau_pl_2}
\end{equation}
As follows from Table~\ref{table:current_prob-2}, for the PE current density considered in this study,
$\tau_{\rm pl}^{(1)} = 14.5~\mu$s for a 200-keV PE and $\tau_{\rm pl}^{(1)} = 1.8~\mu$s for a 30-eV SE.
Substituting these values into Eqs.~(\ref{eq:tau_pl_2-1})--(\ref{eq:tau_pl_2}) and considering a characteristic time for the relaxation of a plasmon excitation $\tau_{\rm rel} \sim 10^2$~ps,
one obtains:
\begin{equation}
\tau_{\rm pl}^{(2)} = 14.5~\mu{\rm s} \times \frac{14.5~\mu{\rm s}}{100~{\rm ps}} \approx 2.1~\rm{s}
\end{equation}
for a 200-keV PE and
\begin{equation}
\tau_{\rm pl}^{(2)} = 1.8~\mu{\rm s} \times \frac{1.8~\mu{\rm s}}{100~{\rm ps}} \approx 0.03~\rm{s} \ .
\end{equation}
for a 30-eV SE.
Thus, for the low-energy SEs, about 25 such events will take place, on average, during the acquisition time for one STEM frame from Ref.~\cite{Wang_2012_PRL.108.245502}.
In this case, the energy transfer to the cluster is equal to
\begin{equation}
\Delta \varepsilon_{\rm pl}^{(2)} = 2 \Delta \varepsilon_{\rm av} \ ,
\label{eq:DeltaE_pl_2}
\end{equation}
where $\Delta \varepsilon_{\rm av}$ is the energy transfer to the deposited Au$_{923}$ cluster due to a single plasmon excitation with the energy transfer $\Delta \varepsilon < I_{\rm p}$, see Eq.~(\ref{eq:DeltaEps_av}).

Relaxation of plasmon excitations due to electron-phonon coupling will lead to an increase in the amplitude of atomic vibrations, which will result in an increase in temperature of the cluster.
The expected temperature increase can be estimated from the relation
\begin{equation}
\Delta \varepsilon_{\rm pl}^{(2)} = \frac32 N k_{\rm B}\Delta T \ ,
\end{equation}
where $N = 923$ is the number of atoms in the cluster, $k_{\rm B}$ is the Boltzmann's constant, and $\Delta \varepsilon_{\rm pl}^{(2)}$ is given by Eq.~(\ref{eq:DeltaE_pl_2}). The estimate gives the temperature increase $\Delta T \sim 45$~K for $\Delta \varepsilon_{\rm av} \sim 2.65$~eV.

Similarly, one can evaluate the probability that a projectile electron inducing a plasmon excitation in the Au$_{923}$ cluster will transfer the amount of energy above $I_{\rm p}$:
\begin{equation}
P_{\rm pl}^{i} \equiv P_{\Delta \varepsilon > I_{\rm p}} =
j \times
\left( { \displaystyle{ \int_{I_{\rm p}}^{E} \frac{{\rm d}\sigma_{\textrm{pl}}}{{\rm d}\Delta\varepsilon} {\rm d} \Delta\varepsilon } } \right) \ .
\label{eq:eq03a}
\end{equation}
The corresponding characteristic appearance time for this event reads as
\begin{equation}
\tau_{\textrm{pl}}^i = \frac{1}{ P_{\rm pl}^{i} } \ .
\end{equation}
The values of $\tau_{\textrm{pl}}^i$ for the case of interactions with a 200-keV PE and a 30-eV SE emitted from a carbon substrate are
listed in Table~\ref{table:current_prob-2}.
For a 200-keV PE, the characteristic occurrence time $\tau_{\textrm{pl}}^i$ for the formation of a plasmon excitation with the excitation energies $\Delta\varepsilon > I_{\rm p}$ (which will result in the ionization of the cluster) is comparable with the time $\tau_{\rm pl}^{(1)}$. In contrast, for a 30-eV SE,
the time $\tau_{\textrm{pl}}^i$ is an order of magnitude longer than the occurrence time for the formation of a plasmon excitation with the excitation energies $\Delta\varepsilon < I_{\rm p}$.

Now let us evaluate the SE flux density $\tilde{j}_{\rm SE}$ caused by the ionization of the deposited cluster by PEs.
The flux density of the electrons emitted from the deposited Au$_{923}$ cluster can be estimated as
\begin{equation}
j_i = \frac{1}{ \tau_{\rm pl}^i \, S_{\rm cl}} \ ,
\label{eq:j_i}
\end{equation}
where $S_{\rm cl} \approx 7.55$~nm$^2$ is the cluster cross-sectional area %(see Sect.~\ref{sec:irradiation_cond})
and
$\tau_{\rm pl}^i = 15.2$~$\mu$s for $j_{\textrm{PE}} = 3 \times 10^6$ nm$^{-2}$s$^{-1}$ and $E_{\rm PE} = 200$~keV,
see Table~\ref{table:current_prob-2}.
Substituting these values into Eq.~(\ref{eq:j_i}) one obtains
\begin{equation}
j_i \sim 8.7 \times 10^3 \frac{e^-}{{\rm nm}^2 \, {\rm s}} \ .
\end{equation}

The flux density of SEs induced by $j_i$ can be evaluated as
\begin{equation}
\frac{ {\rm d}j_{\rm SE}^i }{ {\rm d}E } = N_{\rm SE}(E) \, \frac{ {\rm d}j_i }{ {\rm d}E }(E) \ ,
\end{equation}
where $N_{\rm SE}(E)$ is the number of SEs emitted from the cluster per one PE of energy $E$.
Then $j_{\rm SE}^i$ within the interval of PE energies $\left[ E ; E + \Delta E \right]$ is equal to
\begin{equation}
j_{\rm SE}^i = N_{\rm SE}(E) \, \frac{ {\rm d}j_i }{ {\rm d}E }(E) \, \Delta E \ .
\end{equation}
A detailed energy distribution of SEs has not been elaborated in this study, but one can assume that
the flux density of SEs within the interval $\left[ E ; E + \Delta E \right]$ is comparable
to the flux density $j_i$:
\begin{equation}
\frac{ {\rm d}j_i }{ {\rm d}E }(E) \, \Delta E \sim \lambda  \, j_i \ ,
\end{equation}
where $\lambda < 1$ is a coefficient.
Then,
\begin{equation}
j_{\rm SE}^i = \lambda \, N_{\rm SE}(E) \, j_i \ .
\end{equation}
The SE yield plotted in Fig.~\ref{fig:SE_yield_carbon} has the maximum value $N_{\rm SE}(E) \sim 1$ at
electron energies $E \sim 500$~eV.
The $N_{\rm SE}(E)$ distribution for a gold target has a similar profile with the maximum value $N_{\rm SE}(E) \sim 1.5$ \cite{Lin_Joy_2005}.
Therefore,
\begin{equation}
j_{\rm SE}^i \sim 1.5 \times \left( 8.7 \times 10^3 \frac{e^-}{{\rm nm}^2 \, {\rm s}} \right)
\sim 1.3 \times 10^4 \frac{e^-}{{\rm nm}^2 \, {\rm s}} \ .
\label{eq:j_i_final}
\end{equation}
This number is of the same order of magnitude as the flux density $j_{\rm SE}$ of SEs emitted from the substrate due to the irradiation by PEs, see Eq.~(\ref{eq:j_SE_substrate}).
As follows from this estimate, the resulting flux density of SEs targeting the cluster (i.e. the sum $j_{\rm SE} + j_{\rm SE}^i$) is $\sim$20\% higher than the value given by Eq.~(\ref{eq:j_SE_substrate}).
Therefore, the characteristic times for the occurrence of plasmon excitations should decrease.
The second row in Table~\ref{table:current_prob-2} summarizes the times $\tau_{\textrm{pl}}^{(1)}$ and $\tau_{\textrm{pl}}^i$ for the occurrence of plasmon excitations accounting for the aforementioned correction.

The main conclusion from the analysis carried out in this section is that characteristic occurrence times for plasmon-induced energy relaxation events in deposited gold clusters are on the microsecond timescale.

%%%%%%%%%%%%%%%%%%%%%%%%%%%%%%%%%%%%%%%%%%%%%%%%%%%%%%%%%%%%%%%%%%%%%
%%%%%%%%%%%%%%%%%%%%%%%%%%%%%%%%%%%%%%%%%%%%%%%%%%%%%%%%%%%%%%%%%%%%%
\section{Momentum transfer by primary electrons}
\label{sec:momentum_transfer}

A high-energy PE that elastically scatters from atoms of a deposited gold cluster can transfer momentum to the cluster atoms (without excitation of the electronic subsystem of the cluster) and thus initiate the experimentally observed structural transformations of the clusters. In this section, the probability for the occurrence of such events is evaluated and compared to the probabilities determined in Section~\ref{sec:plasmons}.

The elastic scattering cross section reads as
\begin{equation}
\sigma_{\textrm{el}}
= \int {\frac {\mathrm{d} \sigma_{\textrm{el}} }{\mathrm {d} \Omega }}(\theta ) \, \mathrm {d} \Omega
= 2\pi \int_0^{\pi} {\frac {\mathrm{d} \sigma_{\textrm{el}} }{\mathrm {d} \Omega }}(\theta ) \, \sin \theta \, \mathrm {d} \theta \ ,
\label{eq:MT_CS}
\end{equation}
where ${\mathrm{d} \sigma_{\textrm{el}}/\mathrm {d} \Omega}$ is the differential cross section for elastic scattering, $\Omega$ is a solid scattering angle and $\theta$ is a polar scattering angle.

\begin{figure}[t!]
\centering
\includegraphics[width=0.43\textwidth]{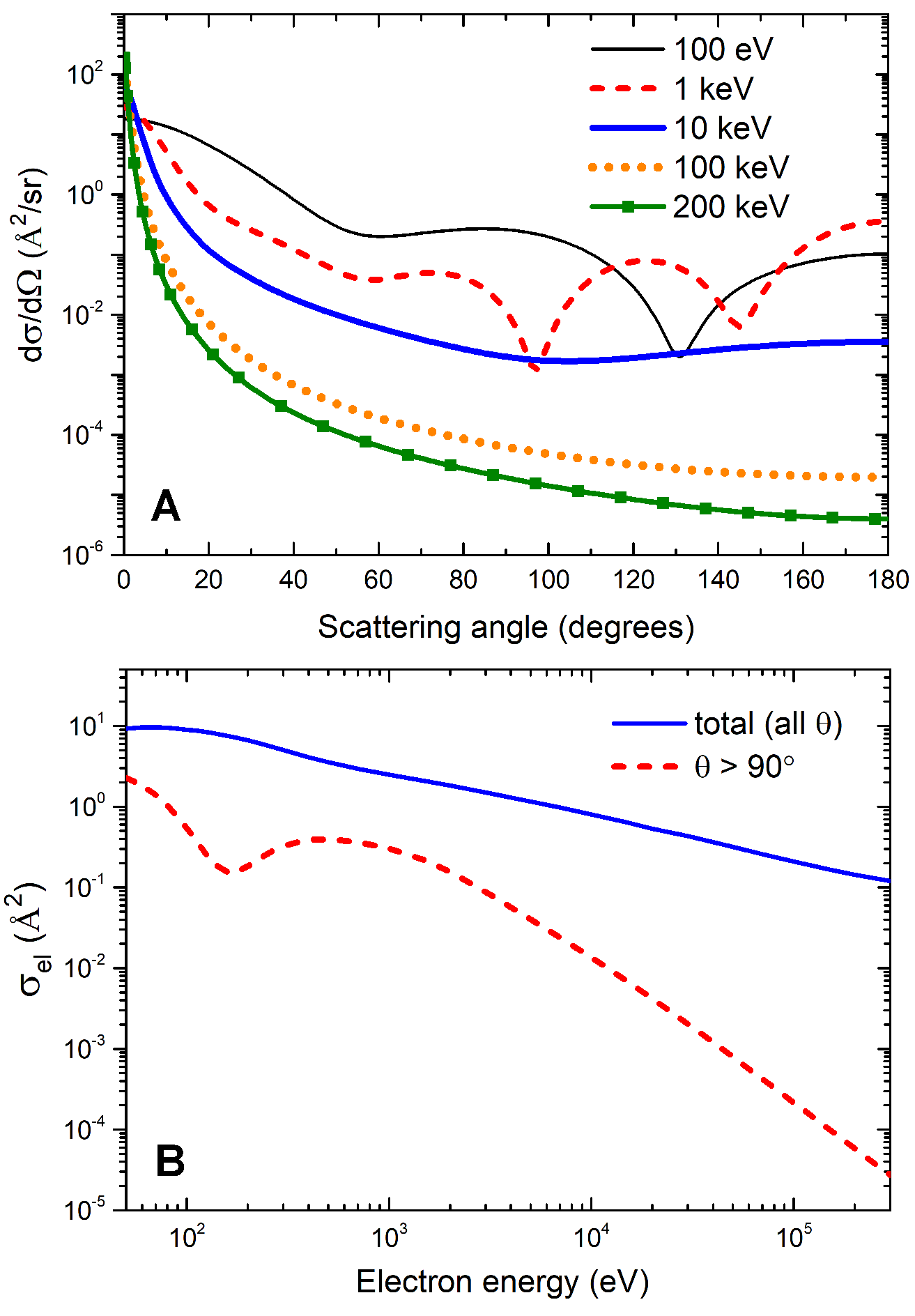}
\caption{Panel~\textbf{A}: the differential cross section ${\mathrm{d} \sigma_{\textrm{el}}/\mathrm {d} \Omega}$ for elastic scattering of an electron from a gold atom for different kinetic energies of a projectile electron. Panel~\textbf{B}: the integral cross section $\sigma_{\textrm{el}}$ calculated using Eq.~(\ref{eq:MT_CS}). The solid line shows the total cross section $\sigma_{\textrm{el}}$ accounting for all possible values of the scattering angle $\theta$. The dashed line shows the partial contribution to $\sigma_{\textrm{el}}$ from electrons scattered in the backward direction ($\theta > 90^{\circ}$). }
\label{fig:MT_CS}
\end{figure}

Figure~\ref{fig:MT_CS}A shows the cross section ${\mathrm{d} \sigma_{\textrm{el}}/\mathrm {d} \Omega}$ for elastic scattering of an electron with the kinetic energy $E$ from a gold atom. The plotted data have been taken from the NIST Electron Elastic-Scattering Cross-Section Database \cite{NIST_elastic}.
Figure~\ref{fig:MT_CS}B shows the integral elastic scattering cross section $\sigma_{\textrm{el}}$ calculated using Eq.~(\ref{eq:MT_CS}). The solid line shows the total cross section $\sigma_{\textrm{el}}$ which accounts for all possible values of the scattering angle $\theta$.
The dashed line shows the partial cross section for electrons scattered in the backward direction ($\theta > 90^{\circ}$), which corresponds to a large value of the momentum transferred to a target atom.

\begin{figure}[t!]
\centering
\includegraphics[width=0.45\textwidth]{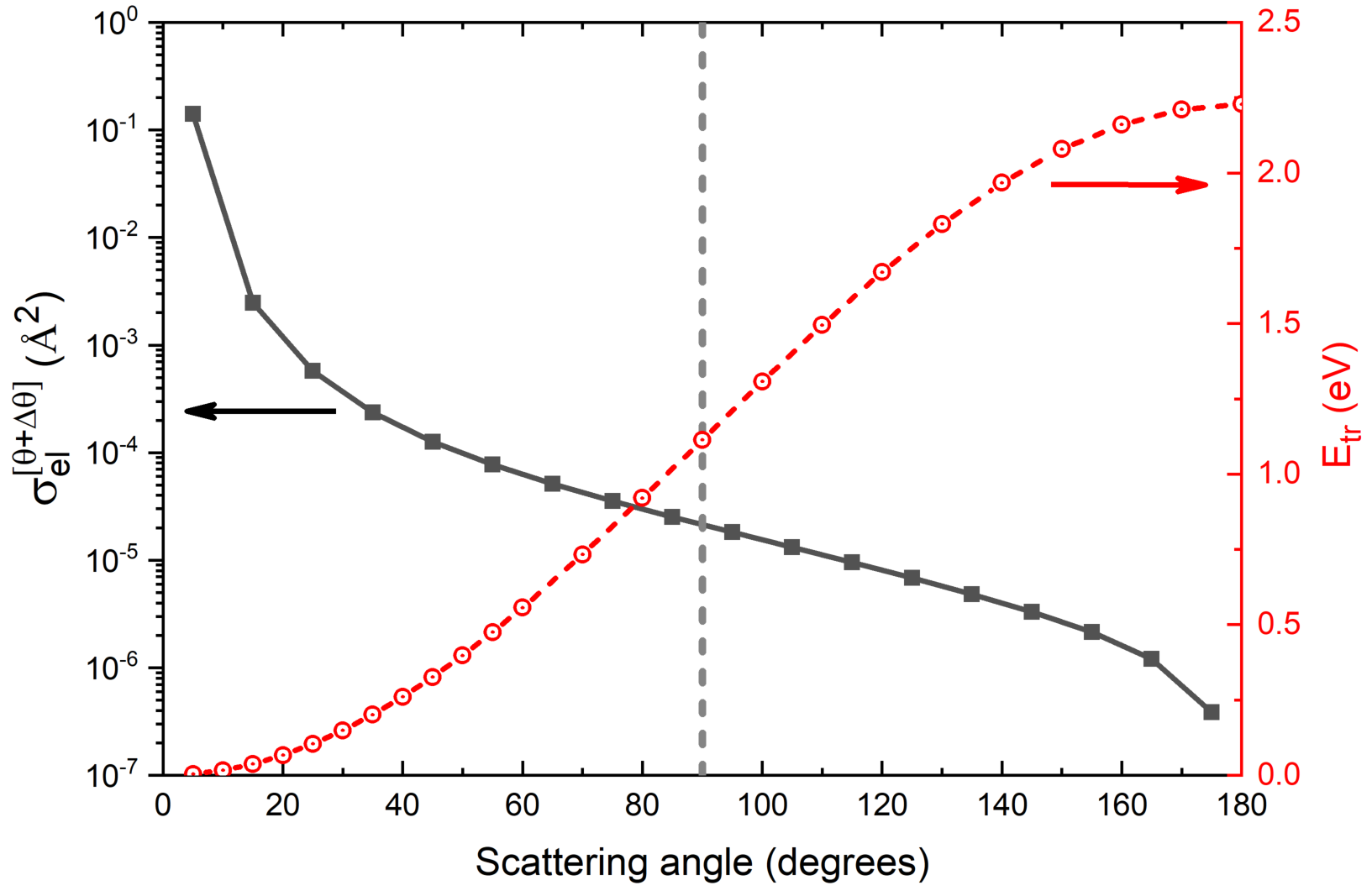}
\caption{The partial cross section $\sigma_{\textrm{el}}^{\left[ \theta; \theta + \Delta\theta \right] }$ for elastic scattering within the interval of scattering angles $\left[ \theta; \theta + \Delta \theta \right]$ (black curve with closed symbols). The red curve with open symbols shows the energy $E_{\rm tr}$ transferred to a gold atom by a 200-keV electron scattered at angle $\theta$, see Eq. (\ref{eq:elastic_collision_E}). }
\label{fig:scattering_energy}
\end{figure}

Let us analyze the partial cross section for elastic scattering within the interval of scattering angles
$\left[ \theta; \theta + \Delta \theta \right]$:
\begin{equation}
\sigma_{\textrm{el}}^{\left[ \theta; \theta + \Delta\theta \right] }
= 2\pi \int_{\theta}^{\theta + \Delta\theta} {\frac {\mathrm{d} \sigma_{\textrm{el}} }{\mathrm {d} \Omega }}(\theta ) \, \sin \theta \, \mathrm {d} \theta \ .
\label{eq:MT_CS-11}
\end{equation}
The dependence of the cross section $\sigma_{\textrm{el}}^{\left[ \theta; \theta + \Delta\theta \right] }$ on $\theta$
is shown in Fig.~\ref{fig:scattering_energy} by the solid black curve.
We have considered the whole range of scattering angles with the bin size $\Delta \theta = 10^{\circ}$.
The probability (per unit time) of electron elastic scattering from the Au$_{923}$ cluster within the interval $\left[ \theta; \theta + \Delta\theta \right]$ is given by:
\begin{equation}
P_{\textrm{el}} =
j_{\textrm{PE}} \, N \, \sigma_{\textrm{el}}^{\left[ \theta; \theta + \Delta\theta \right] } \ ,
\label{eq:eq04a}
\end{equation}
where $N$ is the number of atoms in the cluster.

Due to a rapid decrease of the cross section $\sigma_{\textrm{el}}^{\left[ \theta; \theta + \Delta\theta \right] }$ with increasing $\theta$, the probability $P_{\textrm{el}}$ for electron scattering within the interval $\theta = 170^{\circ}-180^{\circ}$ (corresponding to the largest momentum transfer) is $\sim$50 times lower than the probability for scattering at $\theta = 90^{\circ}-100^{\circ}$ and about five orders of magnitude smaller than that for scattering at small angles $\theta = 0^{\circ}-10^{\circ}$. Each ``soft'' collision will lead to the transfer of a small amount of energy, see the dashed red curve in Fig.~\ref{fig:scattering_energy} and the discussion below. On the other hand, a projectile electron may experience multiple scattering events at small angles colliding successively with several atoms of the cluster. In this case, the amount of energy transferred to the cluster will be comparable to or even smaller than the amount of energy transferred to the cluster during one ``hard'' collision with a single gold atom. In what follows we focus on the collisions corresponding to the scattering angles $\theta > 90^{\circ}$. A detailed analysis of momentum and energy transfer to the cluster as a result of multiple ``soft'' scattering events might be a subject for a future investigation.

Substituting the experimental value $j_{\textrm{PE}} = 3 \times 10^{6}$ nm$^{-2}$s$^{-1}$ \cite{Wang_2012_PRL.108.245502} into Eq.~(\ref{eq:eq04a}) one obtains
\begin{eqnarray}
P_{\textrm{el}}^{\left[ 90^{\circ}; 100^{\circ} \right] }
&\approx& 5.03 \times 10^{-7}~\textrm{ns}^{-1} \ , \nonumber \\
P_{\textrm{el}}^{\left[ 170^{\circ}; 180^{\circ} \right] }
&\approx& 1.07 \times 10^{-8}~\textrm{ns}^{-1} \ .
\label{eq:eq04}
\end{eqnarray}
The corresponding occurrence times for an electron collision involving a large momentum transfer,
\begin{equation}
\tau_{\textrm{el}} = P_{\textrm{el}}^{-1} \ , % \approx 6.18 \times 10^5~\textrm{ns} \approx 0.618~\textrm{ms} \ .
\end{equation}
vary from $\sim$2~ms for the scattering angle interval $\theta = 90^{\circ}-100^{\circ}$ to 93.4~ms for the interval $\theta = 170^{\circ}-180^{\circ}$.
Thus, $\sim 10 - 400$ such events should happen at the experimental conditions of Ref.~\cite{Wang_2012_PRL.108.245502} over the acquisition time for one STEM frame (equal to 0.8~s).

The evaluated characteristic occurrence times $\tau_{\textrm{el}}$ for large momentum transfer events are $3-4$ orders of magnitude longer than the occurrence times $\tau_{\textrm{pl}}$ for energy transfer into the deposited Au$_{923}$ cluster upon inducing a plasmon excitation (see Table~\ref{table:current_prob-2}).
The relaxation of plasmon excitations due to electron-phonon coupling is therefore a more probable mechanism of experimentally observed structural transformations in deposited clusters compared to large momentum transfer in an electron--atom collision.

The maximum energy transferred to the target atom as a result of the head-on collision ($\theta = 180^{\circ}$) of an electron with the nucleus is given by \cite{Landau_1}
\begin{equation}
E_{\rm tr}^{\rm max} = \frac{4 m_e M}{(m_e + M)^2} E \ ,
\label{eq:elastic_collision_Emax}
\end{equation}
where $m_e$ is the mass of a projectile electron, $E$ is its kinetic energy, and $M$ is the mass of a gold atom.
According to Eq.~(\ref{eq:elastic_collision_Emax}), a gold atom hit by a 200-keV electron will acquire the maximum kinetic energy $E_{\rm tr}^{\rm max} \sim 2.23$~eV.
If one accounts for the relativistic kinematics of a collision between an energetic electron and an atom, Eq.~(\ref{eq:elastic_collision_Emax}) transforms into
\begin{equation}
E_{\rm tr}^{\rm max} = \frac{2 (\gamma + 1) m_e M}{m_e^2 + M^2 + 2\gamma m_e M} E \ ,
\label{eq:elastic_collision_Emax_REL}
\end{equation}
where $\gamma = (1 - (v/c)^2)^{-1/2}$ is the Lorentz factor and $c$ is the speed of light in vacuum. For a 200-keV electron (with the speed $v \approx 0.69c$) colliding with a gold atom, the relativistic maximum energy transfer is $E_{\rm tr}^{\rm max} \approx 2.66$~eV. A detailed analysis of the energy transfer processes with accounting for the relativistic kinematics is a separate research question which can be addressed in future studies.

The energy transferred to an atom by an electron scattered at angle $\theta$ is given by \cite{Landau_1}:
\begin{equation}
E_{\rm tr}(\theta) = E_{\rm tr}^{\rm max} \, \sin^2 \frac{\theta}{2} \ .
\label{eq:elastic_collision_E}
\end{equation}
The dependence $E_{\rm tr}(\theta)$ is shown in Fig.~\ref{fig:scattering_energy} by the dashed red curve.
The average amount of energy transferred to a single gold atom during the collision at $\theta < 90^{\circ}$ is $\langle E_{\rm tr} \rangle \sim 0.01$~eV, which is an order of magnitude smaller than the average amount of energy transferred to a single gold atom at $\theta > 90^{\circ}$, $\langle E_{\rm tr} \rangle \sim 1.5$~eV.

%%%%%%%%%%%%%%%%%%%%%%%%%%%%%%%%%%%%%%%%%%%%%%%%%%%%%%%%%%%%%%%%%%%%%
%%%%%%%%%%%%%%%%%%%%%%%%%%%%%%%%%%%%%%%%%%%%%%%%%%%%%%%%%%%%%%%%%%%%%
\section{MD simulations of structural transformations in deposited clusters}
\label{sec:MD_simulations}

%%%%%%%%%%%%%%%%%%%%%%%%%
\begin{figure}[b!]
\centering
\includegraphics[width=0.38\textwidth]{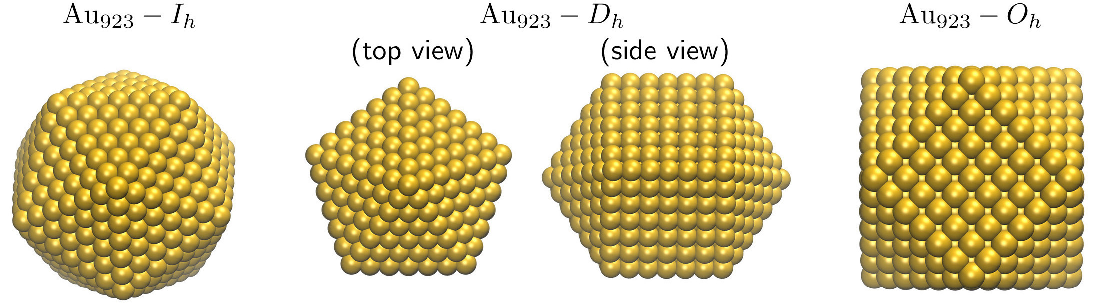}
\caption{Atomistic structures of icosahedron ($I_h$), decahedron ($D_h$) and fcc / octahedron ($O_h$) isomers of the Au$_{923}$ cluster.}
\label{fig:Au923_isomers}
\end{figure}
%%%%%%%%%%%%%%%%%%%%%%%%%

%%%%%%%%%%%%%%%%%%%%%%%%%
\begin{figure*}[t!]
\centering
\includegraphics[width=0.75\textwidth]{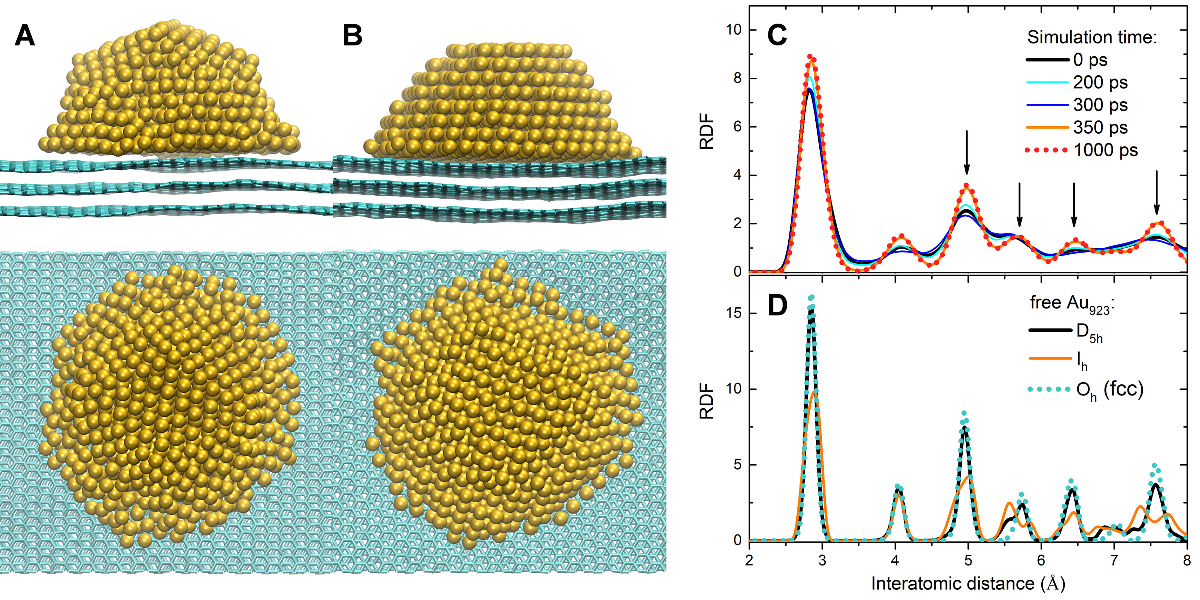}
\caption{Panel~\textbf{A}: the side and top views of the icosahedron-like Au$_{923}$ cluster softly deposited on graphite. Panel~\textbf{B} shows the final geometry of the cluster after the structural transformation to an fcc structure induced by the vibrational excitation of the cluster. Panel~\textbf{C}: Radial distribution functions (RDFs) for the deposited Au$_{923}$ cluster after the energy $\Delta \varepsilon = 2.65$~eV has been uniformly deposited into the cluster. The chosen value of deposited energy corresponds to a characteristic energy of a plasmon excitation which will decay with a significant probability through the vibrational excitation of the ionic subsystem. Arrows show the appearance of peaks in the RDFs indicative for a $I_h \to {\rm fcc}$ structural transformation. Reference RDFs for the free decahedron ($D_{h}$), icosahedron ($I_h$) and fcc/octahedron ($O_h$) Au$_{923}$ clusters are plotted for comparison in Panel~\textbf{D}.}
\label{fig:Au923_RDF}
\end{figure*}
%%%%%%%%%%%%%%%%%%%%%%%%%

The theoretical analysis carried out in Sections~\ref{sec:plasmons} and \ref{sec:momentum_transfer} has been complemented by classical molecular dynamics (MD) simulations  performed by means of MBN Explorer \cite{MBNExplorer_JCC_2012} and MBN Studio \cite{MBNStudio_paper_2019} software packages.

First, decahedron ($D_{h}$), icosahedron ($I_h$) and fcc/cubic ($O_h$) isomers of Au$_{923}$ have been created using the Atomistic Simulation Environment tool \cite{ASE_paper}, see Fig.~\ref{fig:Au923_isomers}.
The interaction between gold atoms has been described using the Gupta potential \cite{Gupta_1983_PRB.23.6265} with the parameters taken from Ref.~\cite{cleri1993tight} and the interaction cutoff of 7~\AA.
Each cluster geometry has been optimized using the velocity quenching algorithm.
The calculated potential energy of a free $D_{h}$ isomer is lower by $\sim$0.94~eV than the energies of free $I_h$ and $O_h$ isomers.
The energy difference between the optimized geometries of the clusters placed on a graphite substrate decreases to $\sim$0.47~eV, with $D_{h}$ still being the lowest-energy isomer among the studied systems.

In this study, we have focused on simulations of the electron-irradiation induced structural transformations of icosahedral Au$_{923}$ clusters.
The clusters were softly deposited onto a carbon substrate following the procedure described in our earlier study  \cite{Verkhovtsev_2020_EPJD.74.205}. A graphite substrate made of three carbon layers has been considered.
The interaction between gold and carbon atoms was described using the Morse potential with the parameters taken from Ref.~\cite{Verkhovtsev_2020_EPJD.74.205}. After the soft landing, the system has been equilibrated at 300~K for 1~ns using the Langevin thermostat. The equilibrated cluster geometries were used to carry out two sets of simulations.

In the first set of simulations, the energy $\Delta \varepsilon = 2.65$~eV has been uniformly deposited into the cluster, and the evolution of the system's structure was monitored over 1~ns.
The chosen value of $\Delta \varepsilon$ corresponds to a characteristic energy of a plasmon excitation
which will decay with a significant probability through the vibrational excitation of the ionic subsystem, see Sect.~\ref{sec:plasmons}.
The performed simulations correspond to a conservative scenario when the energy $\Delta \varepsilon$ is transferred to the cluster as a result of a single plasmon excitation. As discussed in Sect.~\ref{sec:plasmons}, the amount of energy transferred to the cluster will be twice larger if the second plasmon is excited in the cluster within the period of relaxation of the first plasmon.

In the second set of simulations, the process of energy transfer to specific cluster atoms as a result of elastic collisions with the PEs has been studied. At each simulation step, one gold atom was randomly selected and its velocity was increased according to the excess kinetic energy $E_{\rm tr}$, given by Eq.~(\ref{eq:elastic_collision_E}).
Two limiting values of $E_{\rm tr}$ have been considered: $E_{\rm tr} \approx 1.1$~eV and 2.2~eV corresponding to the scattering angles
$\theta = 90^{\circ}$ and 180$^{\circ}$, respectively.
Then, the system was evolved over 10~ps without a thermostat, enabling the energy given to a specific gold atom to be redistributed between other degrees of freedom of the cluster as well as between the cluster and the substrate. 300 subsequent simulations have been carried out for each $E_{\rm tr}$ value.
The number of ``hard'' head-on collisions of a similar order of magnitude will take place over the characteristic experimental irradiation times of $100-400$~s \cite{Wang_2012_PRL.108.245502}.

%%%%%%%%%%%%%%%%%%%%%%%%%
\begin{figure}[t!]
\centering
\includegraphics[width=0.44\textwidth]{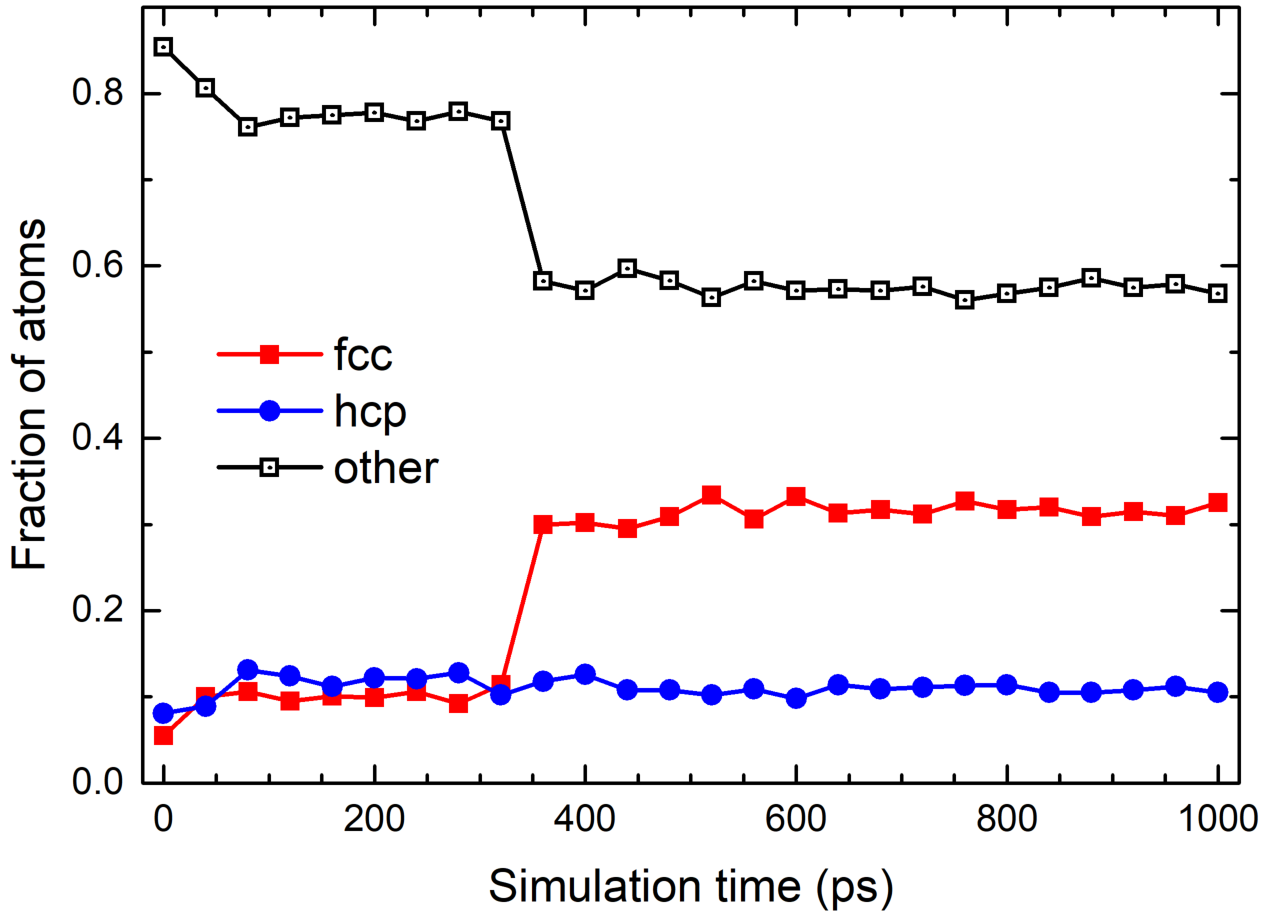}
\caption{Structural analysis of the Au$_{923}$ cluster after the energy $\Delta \varepsilon = 2.65$~eV has been uniformly deposited into the cluster. The figure shows the fraction of atoms in the cluster, which are assigned to a specific crystalline lattice by the CNA algorithm. A rapid increase in the fcc fraction after $\sim$300~ps of the simulation time indicates a $I_h \to {\rm fcc}$ structural transformation induced by the vibrational excitation of the cluster. }
\label{fig:Au923_CNA_plasmon}
\end{figure}
%%%%%%%%%%%%%%%%%%%%%%%%%

Results of the analysis of structural transformations induced by the vibrational excitation of the cluster are summarized in Fig.~\ref{fig:Au923_RDF}. In the course of the simulations the Au$_{923}$--$I_h$ cluster softly deposited on graphite (Fig.~\ref{fig:Au923_RDF}A) has undergone a structural transformation to an fcc structure, see Fig.~\ref{fig:Au923_RDF}B.
This transformation has been quantified by the analysis of radial distribution functions (RDFs) for the initial and final cluster structures, see Fig.~\ref{fig:Au923_RDF}C. Arrows in Fig.~\ref{fig:Au923_RDF}C indicate the appearance of several peaks in the RDF, which are absent in the icosahedral cluster and thus indicative for an $I_h \to {\rm fcc}$ structural transformation.
For comparison, Fig.~\ref{fig:Au923_RDF}D shows reference RDFs for free $D_{5h}$, $I_h$ and $O_h$/fcc Au$_{923}$ clusters.

%%%%%%%%%%%%%%%%%%%%%%%%%
\begin{figure}[t!]
\centering
\includegraphics[width=0.43\textwidth]{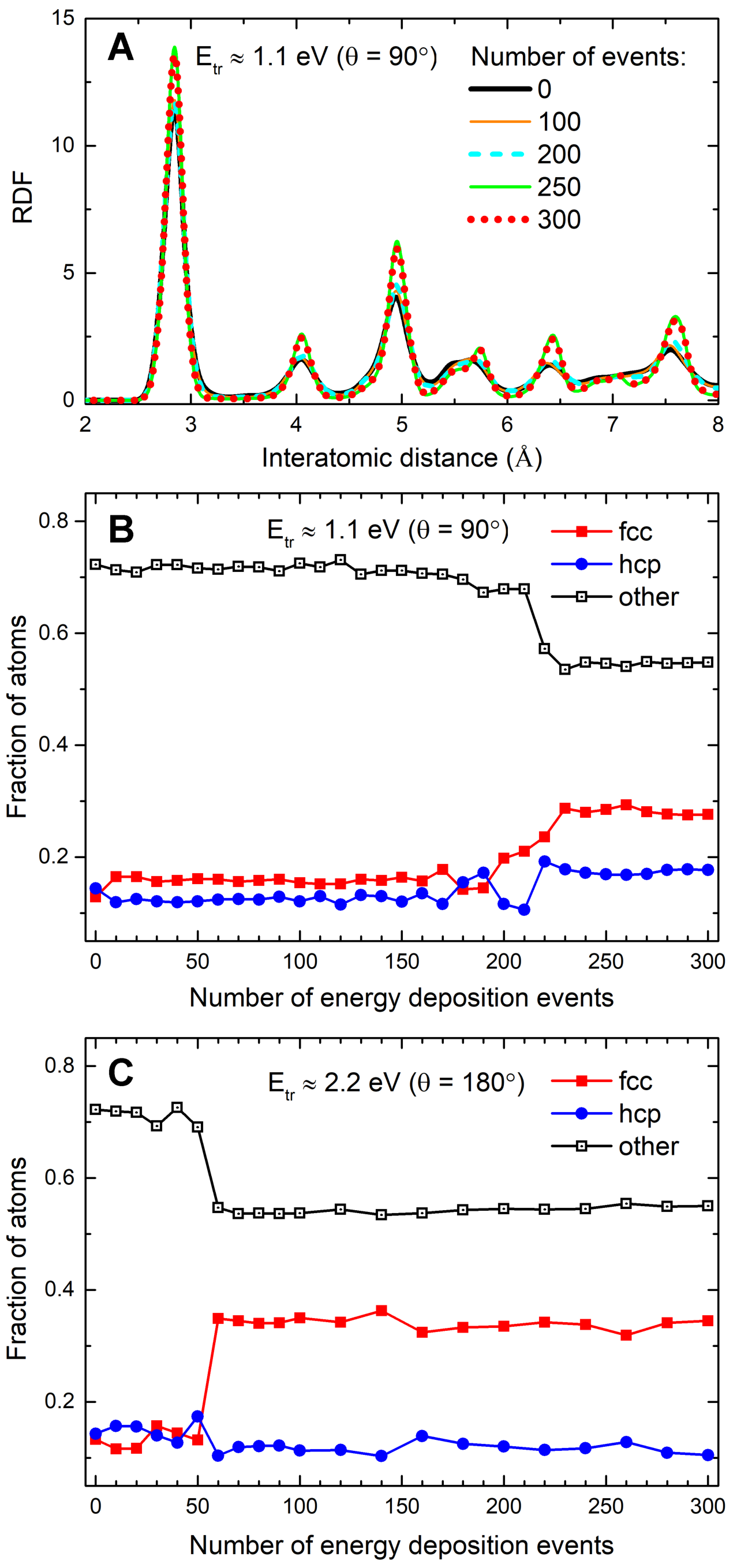}
\caption{Panel~\textbf{A}: RDFs for the deposited Au$_{923}$ cluster after a number of subsequent elastic collisions of the primary 200-keV electrons with atoms of the cluster. In each such collision, the amount of energy $E_{\rm tr} \approx 1.1$~eV corresponding to the scattering angle $\theta = 90^{\circ}$ has been deposited into a randomly selected cluster atom.
Panel~\textbf{B}: Fraction of atoms in the deposited Au$_{923}$ cluster, which are assigned to a specific crystalline lattice by the CNA algorithm. A increase in the fcc fraction from $\sim$15\% to $\sim$30\% indicates a structural transformation in the cluster. This transformation is characterized by the appearance of peaks in the RDFs (shown in panel~A) indicative for an fcc-like structure.
Panel~\textbf{C}: same as panel~B, but for the amount of energy $E_{\rm tr} \approx 2.2$~eV corresponding to the heads-on collision (scattering angle $\theta = 180^{\circ}$). }
\label{fig:Au923_CNA_kicks}
\end{figure}
%%%%%%%%%%%%%%%%%%%%%%%%%

Further details of the observed transformation have been obtained through the structural analysis by means of the common-neighbor analysis (CNA) method \cite{Stukowski_2012_MSMSE.20.045021}, as realized in the OVITO software \cite{Stukowski_2010_MSMSE.18.015012}. Results of this analysis are shown in Fig.~\ref{fig:Au923_CNA_plasmon}. According to the CNA, atoms in the core region of the deposited Au$_{923}$ cluster were arranged initially into the hcp and fcc lattices. Gold atoms in the outer region cannot be assigned by the CNA algorithm to any of the crystalline structures due to their reduced coordination number. Note that the CNA algorithm assigns atoms in the highly-symmetric free Au$_{923}$--$I_h$ cluster (see Fig.~\ref{fig:Au923_isomers}) as belonging to hcp and fcc lattices. Therefore, the distribution of atoms in the fcc and hcp lattices at the beginning of the simulation is indicative of an icosahedral structure.
After $\sim$300~ps of the simulation, a rapid increase in the fcc fraction from $\sim$10\% to $\sim$30\% has occurred, see Fig.~\ref{fig:Au923_CNA_plasmon}. At the same time, the fraction of non-crystalline atoms decreased from $\sim$75\% to $\sim$55\%, indicating that the inner part of the cluster has rearranged into an fcc-like structure.

Results of MD simulations of structural transformations induced by momentum transfer are summarized in Fig.~\ref{fig:Au923_CNA_kicks}.
Panels~A and B illustrate an $I_h \to {\rm fcc}$ structural transformation which has occurred upon the sequential deposition of the energy $E_{\rm tr} \approx 1.1$~eV (corresponding to the scattering angle $\theta = 90^{\circ}$) into randomly selected gold atoms.
Deposition of the energy $E_{\rm tr}$ into different atoms results in the change of RDF, similar to the results shown above in Fig.~\ref{fig:Au923_RDF}.
As shown in Fig.~\ref{fig:Au923_CNA_kicks}B, an increase in the fcc fraction from $\sim$15\% to $\sim$30\% has occurred after ca. 200 energy deposition events, while the fraction of non-crystalline atoms decreased from $\sim$70\% to $\sim$55\%, similar to the results shown in Fig.~\ref{fig:Au923_CNA_plasmon}.
Figure~\ref{fig:Au923_CNA_kicks}C demonstrates that a similar transition takes place upon the deposition of the energy $E_{\rm tr}^{\rm max} \approx 2.2$~eV, corresponding to the scattering angle $\theta = 180^{\circ}$ (heads-on collision). In this case, a smaller number of energy deposition events (about 50 events) are required for the structural transition.

The results shown in Figures~\ref{fig:Au923_CNA_plasmon} and \ref{fig:Au923_CNA_kicks} indicate that the fraction of atoms in the fcc lattice remains nearly constant (for the given simulation times) once the structural transformation has occurred.
This result agrees qualitatively with the experimental observations \cite{Wang_2012_PRL.108.245502} that deposited fcc clusters are more stable than $I_h$ cluster, and no further transformations in Au$_{923}$ have been observed after the $I_h \to D_h$ or $I_h \to \textrm{fcc}$ transformation occurs.
It should be noted that we are unaware of any equivalent analysis of this kind in experiments studying the structure of deposited clusters. Therefore, further efforts have to be made about the detailed comparison of the experimentally observed cluster structures with simulated ones. The important conclusion from the present study is that structural transitions to fcc-like Au$_{923}$ structures are seen both in the experiment \cite{Wang_2012_PRL.108.245502} and the present simulations.

%%%%%%%%%%%%%%%%%%%%%%%%%%%%%%%%%%%%%%%%%%%%%%%%%%%%%%%%%%%%%%%%%%%%%
%%%%%%%%%%%%%%%%%%%%%%%%%%%%%%%%%%%%%%%%%%%%%%%%%%%%%%%%%%%%%%%%%%%%%
\section{Conclusions}

We have presented the results of a theoretical and computational study of structural transformations in deposited nanometer-sized gold clusters exposed to a beam of energetic electrons. The experimentally studied Au$_{923}$ clusters have been considered as an illustrative case study.
The physical mechanisms contributing to the electron-beam induced transformations in deposited metallic clusters have been analyzed and discussed.

We have demonstrated that the relaxation of collective electronic excitations formed in clusters through the vibrational excitation of cluster atoms is a plausible mechanism for the experimentally observed structural transformations.
It has been shown that the characteristic occurrence times for plasmon-induced energy relaxation events are several orders of magnitude shorter than those for the momentum transfer events by energetic primary electrons to atoms of the cluster.
A structural transformation induced by the aforementioned mechanisms has been simulated by means classical molecular dynamics.
The simulations demonstrated that an icosahedral Au$_{923}$ cluster softly deposited on graphite undergoes a structural transformation to an fcc-like structure due to the vibrational excitation of the cluster.

The analysis carried out in this study corresponds to the experimental conditions of Ref.~\cite{Wang_2012_PRL.108.245502}, where deposited Au$_{923}$ clusters were irradiated with a 200-keV electron beam of a scanning transition electron microscope at a beam current $I \approx 53$~pA.
As shown in this paper, both considered mechanisms can contribute to the experimentally observed structural transformations of the deposited gold clusters. The same conclusion is valid for the broad range of primary electron energies typical for STEM experiments, $E_{\rm PE} \sim 30 - 300$~keV.

A possible way to disentangle the contributions of these two mechanisms is to perform experiments similar to those described in Ref.~\cite{Wang_2012_PRL.108.245502} at irradiation regimes when the relaxation of plasmon excitations in deposited gold clusters will be the dominating mechanism of electron irradiation-induced structural transformation compared to heads-on elastic scattering events. These are (i) irradiation with lower-energy keV electrons ($E_{\rm PE} \sim 30-50$~keV) and (ii) irradiation with energetic electrons ($E_{\rm PE} \sim 200-300$~keV) at low beam currents of a few pA.
The maximum energy transfer by electrons with the energies of a few tens of keV is an order of magnitude smaller than that for $200-300$ keV electrons.
Therefore, in low-voltage STEM experiments, significant amounts of energy ($\sim 2-3$~eV) will be transferred to the cluster atoms solely due to the relaxation of collective electronic excitations.
At high-energy irradiation at low beam current (on the order of a few pA), the characteristic occurrence time for the maximum energy transfer due to heads-on elastic collision should be on a few-second scale, which is an order of magnitude longer than the typical acquisition time for one STEM frame reported in Ref.~\cite{Wang_2012_PRL.108.245502}.
A systematic study of the occurrence of structural transformations in clusters of different sizes and for a broader range of irradiation conditions might be addressed in follow-up studies.

%%%%%%%%%%%%%%%%%%%%%%%%%%%%%%%%%%%%%%%%%%%%%%%%%%%%%%%%%%%%%%%%%%%%%
%%%%%%%%%%%%%%%%%%%%%%%%%%%%%%%%%%%%%%%%%%%%%%%%%%%%%%%%%%%%%%%%%%%%%
\section*{Appendix. Expressions for the function $S_l$}
%\appendix
%\section{Expressions for the function $S_l$}
\label{sec:Appendix}

Explicit expressions for the function $S_l$, Eq.~(\ref{eq:funct_S_l}), for different values of $l$ are as follows:
\begin{eqnarray}
S_1(x) &=& \frac{1}{72 x^6} \, \left[ -6 - 9 x^2 + (6 - 3 x^2 + 2 x^4) \cos{(2x)} \right. \nonumber \\
&+& \left. 8 x^6 Ci(2x) + 12x \, \sin{(2x)} \right. \nonumber \\
&+& \left. 2 x^3\, \sin{(2x)} - 4 x^5 \, \sin{(2x)} \right] \ , \label{eq:funct_S_1} % \\
\end{eqnarray}
\begin{eqnarray}
S_2(x) &=& -\frac{1}{16 x^8} \left[ (9 + 4 x^2 + 2 x^4 \right. \nonumber \\
&+& \left. (-9 + 14 x^2) \cos{(2x)} \right. \nonumber \\
&+& \left. 2x (-9 + 2 x^2) \sin{(2x)} \right] \ , \label{eq:funct_S_2}  %\\
\end{eqnarray}
\begin{eqnarray}
S_3(x) &=& \frac{1}{16 x^{10}} \left[ -180 - 45 x^2 - 8 x^4 - 2 x^6 \right. \nonumber \\
&+& \left. (180 - 315 x^2 + 38 x^4) \cos{(2x)} \right. \nonumber \\
&+& \left. 2 x (180 - 75 x^2 + 2 x^4) \sin{(2x)} \right] \ . \label{eq:funct_S_3}
\end{eqnarray}
The function $Ci(x)$ in Eq.~(\ref{eq:funct_S_1}) is the cosine integral,
\begin{equation}
Ci(x)
= - \int_t^{\infty} \frac{ \cos t \, \textrm{d}t}{t}
= \gamma + \ln x + \int_0^x \frac{\cos t - 1}{t} \, \textrm{d}t
\end{equation}
with $\gamma \approx 0.5772$ being the Euler's constant.

%%%%%%%%%%%%%%%%%%%%%%%%%%%%%%%%%%%%%%%%%%%%%%%%%%%%%%%
\begin{acknowledgments}
The authors acknowledge financial support from the Deutsche Forschungsgemeinschaft (Project no. 415716638) and the European Union's Horizon 2020 research and innovation programme – the RADON project (GA 872494) within the H2020-MSCA-RISE-2019 call.
This article is also based upon work from the COST Action CA20129 MultIChem, supported by COST (European Cooperation in Science and Technology).
The possibility of performing computer simulations at the Goethe-HLR cluster of the Frankfurt Center for Scientific Computing is gratefully acknowledged.
\end{acknowledgments}

%%%%%%%%%%%%%%%%%%%%%%%%%%%%%%%%%%%%%%%%%%%%%%%%%%%%%%%%%%%%%%%%%%%%%


\begin{thebibliography}{99}

%%%%%% A. B. LastName, ..., and A. B. FinalName, Article name (plain, no italic), dots at the end of each item
% Book titles - italic and all capitalized

\bibitem{Haberland_clusters_book}
  H. Haberland (ed.),
  \textit{Clusters of Atoms and Molecules: Theory, Experiment, and Clusters of Atoms},
  (Springer-Verlag, Berlin, 1994).


\bibitem{Meiwes-Broer_Clusters_book}
  K.-H. Meiwes-Broer (ed.),
  \textit{Metal Clusters at Surfaces: Structure, Quantum Properties, Physical Chemistry}
  (Springer-Verlag, Berlin, 1999).


\bibitem{LatestAdvances_2008_book}
  J.-P. Connerade and A. V. Solov'yov (eds.),
  \textit{Latest Advances in Atomic Clusters Collision: Fission, Fusion, Electron, Ion and Photon Impact}
  (Imperial College Press, London, 2004).


\bibitem{FrontiersNanoscience_vol12}
  S. T. Bromley and S. M. Woodley (eds.),
  \textit{Computational Modelling of Nanoparticles},
  Frontiers of Nanoscience, vol.~12
  (Elsevier, 2018).


\bibitem{Ekardt_MetalClusters_book}
  W. Ekardt (ed.),
  \textit{Metal Clusters}
  (Wiley, 1999).


\bibitem{Ferrando_book}
  R. Ferrando,
  \textit{Structure and Properties of Nanoalloys}
  (Elsevier, 2016).


\bibitem{Ferrando_RMP_2005}
  F. Baletto and R. Ferrando,
  Structural properties of nanoclusters: Energetic, thermodynamic, and kinetic effects,
  Rev. Mod. Phys. \textbf{77}, 371 (2005).
% 371--423


\bibitem{Barnard_RepProgPhys_2010}
  A. S. Barnard,
  Modelling of nanoparticles: approaches to morphology and evolution,
  Rep. Prog. Phys. \textbf{73}, 086502 (2010).


\bibitem{DySoN_book_Springer_2022}
  I. A. Solov'yov, A. V. Verkhovtsev, A. V. Korol, and A. V. Solov'yov (eds.),
  \textit{Dynamics of Systems on the Nanoscale}
  (Springer International Publishing, Cham, 2022).


\bibitem{Schmidt_1998_Nature.393.238}
  M. Schmidt, R. Kusche, B. von Issendorff, and H. Haberland,
  Irregular variations in the melting point of size-selected atomic clusters,
  Nature \textbf{393}, 238 (1998).
% 238--240


\bibitem{Haberland_2005_PRL.94.035701}
  H. Haberland, T. Hippler, J. Donges, O. Kostko, M. Schmidt, and B. von Issendorff,
  Melting of sodium clusters: where do the magic numbers come from?,
  Phys. Rev. Lett. \textbf{94}, 035701 (2005)


\bibitem{Aguado_2011_AnnuRevPhysChem.62.151}
  A. Aguado and M. F. Jarrold,
  Melting and freezing of metal clusters,
  Annu. Rev. Phys. Chem. \textbf{62}, 151 (2011).
% 151--172


\bibitem{Solov'yov_2005_IJMPB.19.4143}
  A. V. Solov’yov,
  Plasmon excitations in metal clusters and fullerenes,
  Int. J. Mod. Phys. B \textbf{19}, 4143 (2005).
% 4143--4184


\bibitem{Saalmann_1998_PRL.80.3213}
  U. Saalmann and R. Schmidt,
  Excitation and relaxation in atom--cluster collisions,
  Phys. Rev. Lett. \textbf{80}, 3213 (1998).
% 3213--3216


\bibitem{Gerhardt_2003_CPL.3.454}
  P. Gerhardt, M. Niemietz, Y. D. Kim, and G. Gantef\"or,
  Fast electron dynamics in small aluminum clusters: non-magic behavior of a magic cluster,
  Chem. Phys. Lett. \textbf{3-4}, 454 (2003).
% 454--459


\bibitem{Smith_1986_Science.233.872}
  D. J. Smith, A. K. Petford-Long, L. R. Wallenberg, and J. O. Bovin,
  Dynamic atomic-level rearrangements in small gold particles,
  Science \textbf{233}, 872 (1986).
% 872--875


\bibitem{Iijima_1986_PRL.56.616}
  S. Iijima and T. Ichihashi,
  Structural instability of ultrafine particles of metals,
  Phys. Rev. Lett. \textbf{56}, 616 (1986).
% 616--619


\bibitem{Marks_1994_RepProgPhys.57.603}
  L. D. Marks,
  Experimental studies of small particle structures,
  Rep. Prog. Phys. \textbf{57}, 603 (1994).
% 603--649


\bibitem{Li_2008_Nature.451.46}
  Z. Y. Li, N. P. Young, M. Di Vece, S. Palomba, R. E. Palmer, A. L. Bleloch, B. C. Curley, R. L. Johnston, J. Jiang, and J. Yuan,
  Three-dimensional atomic-scale structure of size-selected gold nanoclusters,
  Nature \textbf{451}, 46 (2008).
% 46--48


\bibitem{Wang_2012_PRL.108.245502}
  Z. W. Wang and R. E. Palmer,
  Determination of the ground-state atomic structures of size-selected Au nanoclusters by electron-beam-induced transformation,
  Phys. Rev. Lett. \textbf{108}, 245502 (2012).


\bibitem{Plant_2014_JACS.136.7559}
  S. R. Plant, L. Cao, and R. E. Palmer,
  Atomic structure control of size-selected gold nanoclusters during formation,
  J. Am. Chem. Soc. \textbf{136}, 7559 (2014).
% 7559--7562


\bibitem{Foster_2019_NatComms.10.2583}
  D. M. Foster, Th. Pavloudis, J. Kioseoglou, and R. E. Palmer,
  Atomic-resolution imaging of surface and core melting in individual size-selected Au nanoclusters on carbon,
  Nature Commun. \textbf{10}, 2583 (2019).


\bibitem{Li_2020_ScienceAdv.6.eaay4289}
  Z. Li, H.-Y.T. Chen, K. Schouteden, T. Picot, T.-W. Liao, A. Seliverstov, C. Van Haesendonck, G. Pacchioni, E. Janssens, and P. Lievens,
  Unraveling the atomic structure, ripening behavior, and electronic structure of supported Au$_{20}$ clusters,
  Science Adv. \textbf{6}, eaay4289 (2020).


\bibitem{Wang_2012_Nanoscale.4.4947}
  Z. W. Wang and R. E. Palmer,
  Direct atomic imaging and dynamical fluctuations of the tetrahedral Au$_{20}$ cluster,
  Nanoscale \textbf{4}, 4947 (2012).
% 4947--4949


\bibitem{Wang_2012_NanoLett.12.5510}
  Z. W. Wang and R. E. Palmer,
  Experimental evidence for fluctuating, chiral-type Au$_{55}$ clusters by direct atomic imaging,
  Nano Lett. \textbf{12}, 5510 (2012).
% 5510--5514


\bibitem{Knez_2018_Ultramicroscopy}
  D. Knez, M. Schnedlitz, M. Lasserus, A. Schiffmann, W. E. Ernst, and F. Hofer,
  Modelling electron beam induced dynamics in metallic nanoclusters,
  Ultramicroscopy \textbf{192}, 69 (2018).
% 69--79


\bibitem{Gerchikov2000}
  L. G. Gerchikov, A. N. Ipatov, A. V. Solov'yov, and W. Greiner,
  Non-adiabatic electron-ion coupling in dynamical jellium model for metal clusters,
  J. Phys. B: At. Mol. Opt. Phys. \textbf{33}, 4905 (2000).
% 4905--4926


\bibitem{MBNExplorer_JCC_2012}
  I. A. Solov'yov, A. V. Yakubovich, P. V. Nikolaev, I. Volkovets, and A. V. Solov'yov,
  MesoBioNano Explorer -- A universal program for multiscale computer simulations of complex molecular structure and dynamics,
  J. Comput. Chem. \textbf{33}, 2412 (2012).
% 2412--2439


\bibitem{Egerton_book}
  R. F. Egerton,
  \textit{Electron Energy-Loss Spectroscopy in the Electron Microscope}, 3rd ed.
  (Springer Science+Business Media, New York, 2011).


\bibitem{Jiang_2015_RepProgPhys.79.016501}
  N. Jiang,
  Electron beam damage in oxides: a review,
  Rep. Prog. Phys. \textbf{79}, 016501 (2015).


\bibitem{Egerton_2019_Micron.119.72}
  R. F. Egerton,
  Radiation damage to organic and inorganic specimens in the TEM,
  Micron \textbf{119}, 72 (2019).
% 72--87


\bibitem{Susi_2019_NatRevPhys.1.397}
  T. Susi, J. C. Meyer, and J. Kotakoski,
  Quantifying transmission electron microscopy irradiation effects using two-dimensional materials,
  Nature Rev. Phys. \textbf{1}, 397 (2019).


\bibitem{Egerton_2021}
  R. F. Egerton,
  Radiation damage and nanofabrication in TEM and STEM,
  Microscopy Today \textbf{29}, 56 (2021).
% 56--59


\bibitem{Egerton_2004_Micron.35.399}
  R. F. Egerton, P. Li, and M. Malac,
  Radiation damage in the TEM and SEM,
  Micron \textbf{35}, 399 (2004).
% 399--409


\bibitem{Cretu_2012_Carbon.50.259}
  O. Cretu, J. A. Rodr\'{i}guez-Manzo, A. Demorti\`{e}re, and F. Banhart,
  Electron beam-induced formation and displacement of metal clusters on graphene, carbon nanotubes and amorphous carbon,
  Carbon \textbf{50}, 259 (2012).
% 259--264


\bibitem{FEI_STEM_2005}
  M. van der Stam, M. Stekelenburg, B. Freitag, D. Hubert, and J. Ringnalda,
  A new aberration-corrected transmission electron microscope for a new era,
  Miscrosc. Analysis \textbf{19}, 9 (2005).
% 9--11
% STEM electron energy: 80-300 keV


\bibitem{Sun_JMaterSci_STEM_2020}
  C. Sun, S. Lux, E. M\"uller, M. Meffert, and D. Gerthsen,
  Versatile application of a modern scanning electron microscope for materials characterization,
  J. Mater. Sci. \textbf{55}, 13824 (2020).
%  13824--13835


\bibitem{STEM_SmallMethods_2021}
  F. U. Kosasih, S. Cacovich, G. Divitini, and C. Ducati,
  Nanometric chemical analysis of beam-sensitive materials: A case study of STEM-EDX on perovskite solar cells,
  Small Methods \textbf{5}, 2000835 (2021).


\bibitem{Kreibig_Vollmer_book}
  U. Kreibig and M. Vollmer,
  \textit{Optical Properties of Metal Clusters}
  (Springer-Verlag, Berlin-Heidelberg, 1995).


\bibitem{Cluster_Dynamics_book}
  P.-G. Reinhard and E. Suraud,
  \textit{Introduction to Cluster Dynamics}
  (Wiley, 2004).


\bibitem{Iakoubovskii_2008_PRB}
  K. Iakoubovskii, K. Mitsuishi, Y. Nakayama, and K. Furuya,
  Mean free path of inelastic electron scattering in elemental solids and oxides using transmission electron microscopy: Atomic number dependent oscillatory behavior,
  Phys. Rev. B \textbf{77}, 104102 (2008).


\bibitem{NIST_IMFP_database}
  C. J. Powell and A. Jablonski,
  \textit{NIST Electron Inelastic-Mean-Free-Path Database, Version 1.2}
  (National Institute of Standards and Technology, Gaithersburg, MD, 2010).


\bibitem{Powell_1999}
  C. J. Powell and A. Jablonski,
  Evaluation of calculated and measured electron inelastic mean free paths near solid surfaces,
  J. Phys. Chem. Ref. Data \textbf{28}, 19 (1999).
% 19--62


\bibitem{Lin_Joy_2005}
  Y. Lin and D. C. Joy,
  A new examination of secondary electron yield data,
  Surf. Interface Anal. \textbf{37}, 895 (2005).
% 895--900


\bibitem{Bellissimo_2020_JElSpectr}
  A. Bellissimo, G. M. Pierantozzi, A. Ruocco, G. Stefani, O. Yu. Ridzel, V. Asta\v{s}auskas, W. S. M. Werner, and M. Taborelli,
  Secondary electron generation mechanisms in carbon allotropes at low impact electron energies,
  J. Electron Spectros. Relat. Phenom. \textbf{241}, 146883 (2020).


\bibitem{Neukirch_2016_JPCC.116.15034}
  A. J. Neukirch, Z. Guo, and O. V. Prezhdo,
  Time-domain ab initio study of phonon-induced relaxation of plasmon excitations in a silver quantum dot,
  J. Phys. Chem. C \textbf{116}, 15034 (2012).
% 15034--15040


\bibitem{Wang_1993_CPL.205.521}
  Y. Wang, C. Lewenkopf, D. Tom\'{a}nek, G. Bertsch, and S. Saito,
  Collective electronic excitations and their damping in small alkali clusters,
  Chem. Phys. Lett. \textbf{205}, 521 (1993).
% 521--528


\bibitem{Montag_1995_PRB.51.14686}
  B. Montag and P.-G. Reinhard,
  Width of the plasmon resonance in metal clusters,
  Phys. Rev. B \textbf{51}, 14686 (1995).
% 14686--14692


\bibitem{Stietz_2000_PRL.84.5644}
  F. Stietz, J. Bosbach, T. Wenzel, T. Vartanyan, A. Goldmann, and F. Tr\"ager,
  Decay times of surface plasmon excitation in metal nanoparticles by persistent spectral hole burning,
  Phys. Rev. Lett. \textbf{84}, 5644 (2000).


\bibitem{Kresin_2006_PRB.73.115412}
  V. V. Kresin and Yu.N. Ovchinnikov,
  Fast electronic relaxation in metal nanoclusters via excitation of coherent shape deformations,
  Phys. Rev. B \textbf{73}, 115412 (2006).


\bibitem{Bashevoy_2006_NanoLett.6.1113}
  M. V. Bashevoy, F. Jonsson, A. V. Krasavin, N. I. Zheludev, Y. Chen, and M. I. Stockman,
  Generation of traveling surface plasmon waves by free-electron impact,
  Nano Lett. \textbf{6}, 1113 (2006).
% 1113--1115


\bibitem{Koh_2009_ACSNano.3.3015}
  A. L. Koh, K. Bao, I. Khan, W. E. Smith, G. Kothleitner, P. Nordlander, S. A. Maier, and D. W. McComb,
  Electron energy-loss spectroscopy (EELS) of surface plasmons in single silver nanoparticles and dimers: Influence of beam damage and mapping of dark modes,
  ACS Nano \textbf{3}, 3015 (2009).
% 3015--3022


\bibitem{Rossouw_2011_NanoLett.11.1499}
  D. Rossouw, M. Couillard, J. Vickery, E. Kumacheva, and G. A. Botton,
  Multipolar plasmonic resonances in silver nanowire antennas imaged with a subnanometer electron probe,
  Nano Lett. \textbf{11}, 1499 (2011).
% 1499--1504


\bibitem{Wu_2017_ChemRev.118.2994}
  Y. Wu, G. Li, and J. P. Camden,
  Probing nanoparticle plasmons with electron energy loss spectroscopy,
  Chem. Rev. \textbf{118}, 2994 (2017).
% 2994--3031

%%%%%%%%%%%%%%% NEW REFS %%%%%%%%%%%%%%%

\bibitem{Fujimoto_1968_JPhysSocJpn.25.1679}
  H. F. Fujimoto and K. Komaki,
  Plasma oscillations excited by a fast electron in a metallic particle,
  J. Phys. Soc. Jpn \textbf{25}, 1679 (1968).
% 1679--1687
% 10.1143/JPSJ.25.1679


\bibitem{Lushnikov_1975_ZPhysB.21.357}
  A. A. Lushnikov and A. J. Simonov,
  Excitation of surface plasmons in metal particles by fast electrons and x rays,
  Z. Phys. B \textbf{21}, 357 (1975).
% 357--362


\bibitem{Barberan_1985_PRB.31.6354}
  N. Barber\'{a}n and J. Bausells,
  Plasmon excitation in metallic spheres,
  Phys. Rev. B \textbf{31}, 6354 (1985).
% 6354--6359
% 10.1103/physrevb.31.6354


\bibitem{Ferrell_1987_PRB.35.7365}
  T. L. Ferrell, R. J. Warmack, V. E. Anderson, and P. M. Echenique,
  Analytical calculation of stopping power for isolated small spheres,
  Phys. Rev. B \textbf{35}, 7365 (1987).
% 7365--7371
% 10.1103/physrevb.35.7365 


\bibitem{Michalewicz_1992_PRB.45.13664}
  M. T. Michalewicz,
  Identification of plasmons on small metallic particles,
  Phys. Rev. B \textbf{45}, 13664 (1992).
% 13664--13670
% 10.1103/physrevb.45.13664 


\bibitem{Gerchikov_1997_JPB.30.4133}
  L. G. Gerchikov, A. V. Solov'yov, J.-P. Connerade, and W. Greiner,
  Scattering of electrons on metal clusters and fullerenes,
  J. Phys. B: At. Mol. Opt. Phys. \textbf{30}, 4133 (1997).
% 4133--4161


\bibitem{Gerchikov_2000_PRA.62.043201}
  L. G. Gerchikov, A. N. Ipatov, R. G. Polozkov, and A. V. Solov'yov,
  Surface- and volume-plasmon excitations in electron inelastic scattering on metal clusters,
  Phys. Rev. A 62, 043201 (2000).
% 10.1103/PhysRevA.62.043201


\bibitem{GdeAbajo_2010_PMP.82.209}
  F. J. Garc\'{i}a de Abajo,
  Optical excitations in electron microscopy,
  Rev. Mod. Phys. \textbf{82}, 209 (2010).
% 209--275
% 10.1103/revmodphys.82.209


\bibitem{Gildenburg_2016_PhysPlasmas.23.032120}
  V. B. Gildenburg, V. A. Kostin, and I. A. Pavlichenko,
  Excitation of surface and volume plasmons in a metal nanosphere by fast electrons,
  Phys. Plasmas \textbf{23}, 032120 (2016).
% 10.1063/1.4944395

%%%%%%%%%%%%%%%%%%%%%%%%%%%%%%%%%%%%%%%%

\bibitem{Gerchikov_1998_PRL.81.2707}
  L. G. Gerchikov, P. V. Efimov, V. M. Mikoushkin, and A. V. Solov'yov,
  Diffraction of fast electrons on the fullerene C$_{60}$ molecule,
  Phys. Rev. Lett. \textbf{81}, 2707 (1998).


\bibitem{Verkhovtsev_2015_PRL.114.063401}
  A. V. Verkhovtsev, A. V. Korol, and A. V. Solov'yov,
  Revealing the mechanisms of the low-energy electron yield enhancement from sensitizing nanoparticles,
  Phys. Rev. Lett. \textbf{114}, 063401 (2015).


\bibitem{Verkhovtsev_2015_JPCC.119.11000}
  A. V. Verkhovtsev, A. V. Korol, A. V. Solov'yov,
  Electron production by sensitizing gold nanoparticles irradiated by fast ions,
  J. Phys. Chem. C \textbf{119}, 11000 (2015).
% 11000--11013


\bibitem{Kleinig_1998_EPJD.4.343}
  W. Kleinig, V. O. Nesterenko, P.-G. Reinhard, and Ll. Serra,
  Plasmon response in K, Na and Li clusters: systematics using the separable random-phase-approximation with pseudo-Hamiltonians,
  Eur. Phys. J. D \textbf{4}, 343 (1998).
% 343--352


\bibitem{Hoevel_1993_PRB.48.18178}
  H. H\"ovel, S. Fritz, A. Hilger, U. Kreibig, and M. Vollmer,
  Width of cluster plasmon resonances: Bulk dielectric functions and chemical interface damping,
  Phys. Rev. B \textbf{48}, 18178 (1993).
% 18178--18188


\bibitem{Lushnikov_1974_ZPhysik.270.17}
  A. A. Lushnikov and A. J. Simonov,
  Surface plasmons in small metal particles,
  Z. Physik \textbf{270}, 17 (1974).
% 17--24


\bibitem{Korol_AVS_BrS_2014}
  A. V. Korol and A. V. Solov'yov,
  \textit{Polarization Bremsstrahlung},
  Springer Series on Atomic, Optical, and Plasma Physics, Vol.~80
  (Springer, 2014).


\bibitem{Sakata_2016_JAP.120.244901}
  D. Sakata, S. Incerti, M.C. Bordage, N. Lampe, S. Okada, D. Emfietzoglou, I. Kyriakou, K. Murakami, T. Sasaki, H. Tran, S. Guatelli, and V. N. Ivantchenko,
  An implementation of discrete electron transport models for gold in the Geant4 simulation toolkit,
  J. Appl. Phys. \textbf{120}, 244901 (2016).


%\bibitem{Aeschlimann_2000_ApplPhysA.71.485}
%  M. Aeschlimann, M. Bauer, S. Pawlik, R. Knorren, G. Bouzerar, and K. H. Bennemann,
%  Transport and dynamics of optically excited electrons in metals,
%  Appl. Phys. A, \textbf{71}, 485 (2000).
% 485--491


\bibitem{Seidl_1991_JCP.95.1295}
  M. Seidl, K.-H. Meiwes-Broer, and M. Brack,
  Finite-size effects in ionization potentials and electron affinities of metal clusters,
  J. Chem. Phys. \textbf{95}, 1295 (1991).
% 1295--1303


\bibitem{Seidl_1998_JCP.108.8182}
  M. Seidl, J. P. Perdew, M. Brajczewska, and C. Fiolhais,
  Ionization energy and electron affinity of a metal cluster in the stabilized jellium model: Size effect and charging limit,
  J. Chem. Phys. \textbf{108}, 8182 (1998).
% 8182--8189


\bibitem{Seidl_1996_AnnPhys.245.275}
  M. Seidl and M. Brack,
  Liquid drop model for charged spherical metal clusters,
  Ann. Phys. \textbf{245}, 275 (1996).
% 275--310


\bibitem{CRC_Handbook_Chem-Phys}
  J. Rumble (ed.),
  \textit{CRC Handbook of Chemistry and Physics}, 102th ed.
  (CRC Press, 2021).


\bibitem{NIST_Handbook_Atomic_Data}
  \textit{Handbook of Basic Atomic Spectroscopic Data} (NIST Standard Reference Database 108)
%\\ \url{https://www.nist.gov/pml/handbook-basic-atomic-spectroscopic-data}


\bibitem{Haeberlen_1997_JCP.106.5189}
  O. D. H\"aberlen, S.-C. Chung, M. Stener, and N. R\"osch,
  From clusters to bulk: A relativistic density functional investigation on a series of gold clusters Au$_n$, $n = 6, \dots, 147$,
  J. Chem. Phys. \textbf{106}, 5189 (1997).
% 5189--5201


\bibitem{NIST_elastic}
  NIST Standard Reference Database (SRD) 64 -- \textit{NIST Electron Elastic-Scattering Cross-Section Database}


\bibitem{Landau_1}
  L. D. Landau and E. M. Lifshitz,
  \textit{Mechanics}, Course of Theoretical Physics, Vol.~1, 3rd ed.
  (Butterworth-Heinemann, 1976).


\bibitem{MBNStudio_paper_2019}
  G. B. Sushko, I. A. Solov'yov, and A.V. Solov'yov,
  Modeling MesoBioNano systems with MBN Studio made easy,
  J. Mol. Graph. Model. \textbf{88}, 247 (2019).
% 247--260


\bibitem{ASE_paper}
  A.H. Larsen et al.,
%  A.H. Larsen, J.J. Mortensen, J. Blomqvist, I.E. Castelli, R. Christensen,
%  M. Du{\l}ak, J. Friis, M.N. Groves, B. Hammer, C. Hargus, E.D. Hermes,
%  P.C. Jennings, P.B. Jensen, J. Kermode, J.R. Kitchin, E.L. Kolsbjerg,
%  J. Kubal, K. Kaasbjerg, S. Lysgaard, J.B. Maronsson, T. Maxson,
%  T. Olsen, L. Pastewka, A. Peterson, C. Rostgaard, J. Schi{\o}tz, O. Sch\"utt,
%  M. Strange, K.S. Thygesen, T. Vegge, L. Vilhelmsen, M. Walter, Z. Zeng, K.W. Jacobsen,
  The atomic simulation environment -- a Python library for working with atoms,
  J. Phys.: Condens. Matter \textbf{29}, 273002 (2017).


\bibitem{Gupta_1983_PRB.23.6265}
  R. P. Gupta,
  Lattice relaxation at a metal surface,
  Phys. Rev. B \textbf{23}, 6265 (1981).
% 6265--6270


\bibitem{cleri1993tight}
  F. Cleri and V. Rosato,
  Tight-binding potentials for transition metals and alloys,
  Phys. Rev. B \textbf{48}, 22 (1993).
% 22--33


\bibitem{Verkhovtsev_2020_EPJD.74.205}
  A. V. Verkhovtsev, Y. Erofeev, and A.V. Solov'yov,
  Soft landing of metal clusters on graphite: a molecular dynamics study,
  Eur. Phys. J. D \textbf{74}, 205 (2020).


\bibitem{Stukowski_2012_MSMSE.20.045021}
  A. Stukowski,
  Structure identification methods for atomistic simulations of crystalline materials,
  Modelling Simul. Mater. Sci. Eng. \textbf{20}, 045021 (2012).


\bibitem{Stukowski_2010_MSMSE.18.015012}
  A. Stukowski,
  Visualization and analysis of atomistic simulation data with OVITO -- the Open Visualization Tool,
  Modelling Simul. Mater. Sci. Eng. \textbf{18}, 015012 (2010).

\end{thebibliography}
\end{document}